\documentclass[11pt, a4paper]{article}
\usepackage{jheppub, tensind}

\usepackage{amsmath, amsfonts, amssymb}
\usepackage{slashed}
\usepackage[vcentermath]{youngtab}
\usepackage{empheq}
\usepackage{booktabs}
\usepackage{mathrsfs}
\usepackage{xcolor}
\usepackage{comment}
\usepackage[format=hang,labelfont={bf}]{caption}
\newcommand {\cD}{{\cal D}}

\newcommand {\cH}{{\cal H}}

\newcommand {\cK}{{\cal K}}
\newcommand {\cL}{{\cal L}}

\newcommand {\cR}{{\cal R}}

\newcommand {\cV}{{\cal V}}

\newcommand{\ra}{{\mathrm a}}
\newcommand{\rb}{{\mathrm b}}
\newcommand{\rc}{{\mathrm c}}
\newcommand{\rd}{{\mathrm d}}

\newcommand{\halpha}{{\hat{\alpha}}}
\newcommand{\hbeta}{{\hat{\beta}}}
\newcommand{\hgamma}{{\hat{\gamma}}}

\newcommand{\veps}{\varepsilon}

\newcommand{\eol}{\notag \\}

\newcommand{\pa}{\partial}

\newcommand{\hmu}{{\hat\mu}}
\newcommand{\hnu}{{\hat\nu}}

\newcommand{\g}[1]{\mathsf{#1}}

\newcommand{\GL}{\g{GL}}

\newcommand{\Clif}{\textrm{Clif}}
\newcommand{\ket}[1]{|#1\rangle}
\newcommand{\bra}[1]{\langle #1|}
\newcommand{\braket}[2]{\langle #1 \vert #2 \rangle}
\newcommand{\ol}{\overline}

\newcommand{\rba}{{\ol\ra}}
\newcommand{\rbb}{{\ol\rb}}

\newcommand{\balpha}{{\ol \alpha}}
\newcommand{\bbeta}{{\ol \beta}}

\DeclareMathOperator{\Tr}{Tr}

\makeatletter
\g@addto@macro\bfseries{\boldmath}
\makeatother

\numberwithin{equation}{section}

\title{Notes on Ramond-Ramond spinors and bispinors in double field theory}
\author{Daniel Butter}
\affiliation{George P. and Cynthia W. Mitchell Institute for Fundamental Physics and Astronomy,\\
Texas A\&M University, College Station, TX 77843-4242, USA}
\emailAdd{dbutter@tamu.edu}

\abstract{The Ramond-Ramond sector of double field theory (DFT) can be described either as
an $\g{O}(D,D)$ spinor or an $\g{O}(D-1,1) \times \g{O}(1,D-1)$ bispinor. Both formulations
may be related to the standard polyform expansion in terms of even or odd rank field strengths corresponding to IIA or IIB duality frames. The spinor approach is natural in a (bosonic) metric formulation of DFT, while the bispinor is indispensable for supersymmetric DFT. In these notes, we show how these two approaches may be covariantly connected using a spinorial version of the DFT vielbein, which flattens an $\g{O}(D,D)$ spinor into a bispinor. We also elaborate on details of the bispinor formulation in both even and odd $D$ and elaborate on the distinction between the IIA/IIB/IIA${}^*$/IIB${}^*$ duality frames.}

\begin{document}

\maketitle

%%%%%%%%%%%%%%%%%%%%%%%%%%%%%%%%%%%%%%%%%%%%%%%%%%%%%%%%%%%%%%%%%%%%%%%%%%%%%%%%%
\section{Introduction}
%%%%%%%%%%%%%%%%%%%%%%%%%%%%%%%%%%%%%%%%%%%%%%%%%%%%%%%%%%%%%%%%%%%%%%%%%%%%%%%%%

Double field theory (DFT) \cite{Siegel:1993xq,Siegel:1993th, Hull:2009mi,Hull:2009zb,Hohm:2010jy,Hohm:2010pp} is a reformulation of the massless sector of string theory that incorporates T-duality as a manifest symmetry.\footnote{Duff and Lu \cite{Duff:1989tf,Duff:1990hn} and Tseytlin \cite{Tseytlin:1990nb,Tseytlin:1990va} explored very similar ideas. The prescient work of Siegel \cite{Siegel:1993xq,Siegel:1993th} is equivalent to the frame formulation of double field theory.} It combines the metric $g_{mn}$ and Kalb-Ramond two-form $b_{mn}$ into a single generalized metric $\cH_{MN}$ depending on doubled coordinates $x^M = (x^m, \tilde x_m)$, and generalized diffeomorphisms on these coordinates encompass both conventional diffeomorphisms and $b$-field transformations. Augmented with a generalized dilaton $e^{-2 d} = \sqrt{-g}\, e^{-2 \phi}$, one can construct the universal NS-NS sector in terms of the unique two-derivative action in DFT,
\begin{align}\label{E:DFTaction}
S = \int \rd^D x \, \rd^D \tilde x \, e^{-2 d}\, \mathcal R(\cH)
    = \int \rd^D x\, \rd^D \tilde x\, \sqrt{-g} \,e^{-2 \phi} \Big(
    R + 4 |\nabla \phi|^2 - \frac{1}{12} H_{mnp} H^{mnp}
    \Big)~,
\end{align}
where $\cR(\cH)$ is the curvature scalar of the generalized metric. The action \eqref{E:DFTaction} is invariant under the continuous duality group $\g{O}(D,D; \mathbb R)$, which extends the discrete $\g{O}(d,d; \mathbb Z)$ duality group arising when string theory is compactified on a $d$-torus. The continuous group $\g{O}(D,D; \mathbb R)$ dictates both the structure of generalized diffeomorphisms and the generalized metric.

As in general relativity, one can factorize the generalized metric $\cH_{MN}$ of DFT into the square of a generalized vielbein $V_M{}^A$. In general relativity, the vielbein is an element of the coset $\g{GL}(D) / \g{SO}(D-1,1)$, the denominator being the local Lorentz symmetry that leaves the metric invariant; the vielbein is a local frame field that flattens a curved vector index into a Lorentz vector index. In double field theory, similar statements hold true: the vielbein is an element of $\frac{\g{O}(D,D)}{\g{O}(D-1,1)_L \times \g{O}(1,D-1)_R}$, and it flattens a curved vector index into a pair of $D$-dimensional vectors on which two different Lorentz groups act. The vielbein description is crucial in supergravity, because the gravitino (as well as others fields) transforms as a Lorentz spinor and the vielbein appears explicitly in the supersymmetry transformations; the same is true in supersymmetric double field theory \cite{Hohm:2011nu, Jeon:2011sq, Jeon:2012hp} where both spinors and vector-spinors of $\g{SO}(D-1,1)_L \times \g{SO}(1,D-1)_R$ appear.

Because the massless sector of type II string theories involves also a Ramond-Ramond sector of $p$-forms $C_p$ with $p$ even or odd depending on the duality frame, it is natural to seek a duality-covariant way of incorporating these fields. In fact, the Ramond-Ramond sector was first given a duality covariant description within the $\g{E}_{11}$ formulation of West \cite{West:2003fc,West:2010ev} --- it lies just above the NS-NS fields in the $\g{E}_{11}$ level decomposition \cite{Rocen:2010bk}. In this work, we will remain firmly within the confines of double field theory: but even here there are two alternative (but related) methods. In their approach to type II double field theory \cite{Hohm:2011zr, Hohm:2011dv}, Hohm, Kwak, and Zwiebach described the Ramond-Ramond sector in terms of a (bosonic) spinor of $\g{O}(D,D)$, an idea dating back to \cite{Brace:1998xz, Fukuma:1999jt}. Since this formulation was not supersymmetric, there was no need to introduce the double Lorentz group or to factorize the generalized metric into a generalized vielbein. In contrast, the supersymmetric formulation of Jeon, Lee, Park, and Suh \cite{Jeon:2012hp}, building on earlier work \cite{Jeon:2012kd}, locates the Ramond-Ramond sector in a bispinor of $\g{SO}(D-1,1) \times \g{SO}(1,D-1)$ (for the case $D=10$). This approach is quite natural from the perspective of the supersymmetry transformations.\footnote{A closely related discussion was given in the context of
generalized geometry \cite{Coimbra:2011nw}, where the section condition is explicitly solved.}

These two approaches, while obviously related, have not to our knowledge been explicitly connected in the literature before, at least not without fixing a double Lorentz gauge \cite{Jeon:2012kd}; but it is evident that they ought to be. First, neither of the formulations \cite{Hohm:2011zr, Hohm:2011dv} and \cite{Jeon:2012hp} of type II DFT requires any gauge-fixing, so it would be quite unnatural if no covariant connection were possible. Second, a manifestly supersymmetric formulation of type II DFT would seem to require it. This is because just as the Ramond-Ramond $p$-forms of $\g{GL}(D)$ lift to super $p$-forms of $\g{GL}(D|s)$ in conventional type II superspace (for $D=10$ and $s=32$), spinors of $\g{O}(D,D)$ ought to lift to ``superspinors'' of $\g{OSp}(D,D|2s)$ \cite{Cederwall:2016ukd}. Simultaneously, the DFT supervielbein contains the Ramond-Ramond field strengths in bispinor form \cite{Hatsuda:2014qqa}. This mirrors the situation in conventional type II superspace, where the Ramond-Ramond sector is encoded both in a hierarchy of super $p$-forms and as a bispinor in the superspace torsion constraints (see the appendix of \cite{Wulff:2013kga} for a nice compact discussion).

The goal of this paper is to fill this gap and provide an explicit covariant connection between the spinor and bispinor formalisms. In doing so, we  provide a compact collection of results (some old and some new) pertaining to the Ramond-Ramond sector in both formulations for general $D$. Let us briefly sketch how this works here.

The $\g{O}(D,D)$ spinor description of the RR sector is built upon a Fock space where the gamma matrices are interpreted as raising and lowering operators, $\psi^m = \frac{1}{\sqrt2}\Gamma^m$ and $\psi_m = \frac{1}{\sqrt2} \Gamma_m$  \cite{Hohm:2011zr,Hohm:2011dv}. The Ramond-Ramond polyform $F$ is then identified with a ket $\ket{F}$,
\begin{align}
\ket{F} = \sum_p \frac{1}{p!} F_{m_1 \cdots m_p} \psi^{m_1} \cdots \psi^{m_p} \ket{0}~,
\end{align}
which transforms covariantly under $\g{O}(D,D)$ but is invariant under the double Lorentz group. Whether $F$ is even or odd in $p$ determines whether we lie in a IIA or IIB duality frame. 

The bispinor form arises by introducing the spinorial version of the vielbein $V_A{}^M$. For even $D$, the spinorial vielbein is a bispinor-valued bra $\bra{\slashed{V}}$ obeying
\begin{align}
\bra{\slashed{V}} \Gamma^M = 
    \gamma^\ra \bra{\slashed{V}} V_\ra{}^M +
    \gamma_* \bra{\slashed{V}} \bar\gamma^\rba V_\rba{}^M~.
\end{align}
The explicit construction of this object is given in section \ref{S:BispinorDeven.ExplicitS0}.
The gamma matrices $\gamma^{\ra}$ and $\bar\gamma^\rba$ belong respectively to the left and right Lorentz groups. Then the Ramond-Ramond bispinor $\slashed{\widehat F}$ of \cite{Jeon:2012kd, Jeon:2012hp} can be related to the $\g{O}(D,D)$ spinor via
\begin{align}
\slashed{\widehat F} := e^{d} \braket{\slashed{V}}{F}
    = e^{\phi} \sum_p \frac{1}{p!} \widehat F_{\ra_1 \cdots \ra_p} 
    \gamma^{\ra_1} \cdots \gamma^{\ra_p} \slashed{Z}
\end{align}
where $\widehat F_{\ra_1 \cdots \ra_p}$ are the components of the polyform $e^{-b} F$ flattened with the left-handed vielbein $e_m{}^\ra$. This is in the basis where $V_M{}^A$ is parametrized by $b_{mn}$, $e_m{}^\ra$ and $\bar e_m{}^\rba$, with the two vielbeins (which give the same metric) transforming separately under the left and right Lorentz groups, with the Kalb-Ramond two-form invariant \cite{Jeon:2012kd}. The DFT and supergravity dilatons are related by $e^{-2d} = e^{-2 \phi} \sqrt{-g}$. The bispinor $\slashed{Z}$ is a covariant vacuum bispinor, transforming on the left (right) under the left (right) local Lorentz group. The chirality of $\slashed{\widehat F}$ is fixed, and then the even/odd degree $p$ is related to the chirality of $\slashed{Z}$ and this determines whether we lie in a IIA or IIB duality frame.

When $D$ is odd, the dimension of a bispinor is $2^{(D-1)/2} \times 2^{(D-1)/2}$, and one must introduce an additional two-dimensional space to match the dimension ($2^D$) of a spinor of $\g{O}(D,D)$. The spinorial vielbein is then a \emph{doublet} of bispinor-valued bras $\bra{\slashed{V}_i}$ with $i=1,2$. It is defined by
\begin{align}
\bra{\slashed{V}_i} \Gamma^M = 
    (\sigma_1)_{ij}\, \gamma^\ra \bra{\slashed{V}_j} V_\ra{}^M +
    (\sigma_2)_{ij} \,\bra{\slashed{V}_j} \bar\gamma^\rba V_\rba{}^M
\end{align}
in terms of the Pauli matrices on the additional doublet space. 
The explicit construction of this object is given in section \ref{S:BispinorDodd.ExplicitS0}.
One can then define the RR bispinor $\slashed{\widehat F}_i$ by
\begin{align}
\slashed{\widehat F}_i := e^{d} \braket{\slashed{V}_i}{F}
    = e^{\phi} \sum_p \frac{1}{p!} \widehat F_{\ra_1 \cdots \ra_p} 
    \gamma^{\ra_1} \cdots \gamma^{\ra_p} (\sigma_1^p)_{i j} \slashed{Z}_j
\end{align}
in terms of a covariant vacuum bispinor doublet $\slashed{Z}_i$.

The paper is arranged as follows. We begin in section \ref{S:OddSpinors} with a discussion of $\g{O}(D,D)$ spinors and their respective $\g{Pin}$ and $\g{Spin}$ groups, largely following \cite{Hohm:2011zr,Hohm:2011dv}. In section \ref{S:FlatStuff}, we introduce the spinorial vielbein for arbitrary dimension $D$ and describe the connection between ``curved'' and ``flat'' $\g{O}(D,D)$ spinors. We continue in section \ref{S:TypeIIsugras} with a detailed discussion of how to classify the different type II duality frames in terms of the vielbein $V_M{}^A$ and explain how this relates to the covariant vacuum. In both these sections, we treat the even and odd $D$ cases simultaneously; this is possible by avoiding specifying how a ``flat'' spinor of $\g{O}(D,D)$ decomposes under the local double Lorentz group. In sections \ref{S:BispinorDeven} and \ref{S:BispinorDodd}, we discuss the even and odd dimensional cases in detail. In section \ref{S:SelfDuality}, the various possible self-duality conditions are addressed, and we briefly conclude in section \ref{S:Conclusion}.

%%%%%%%%%%%%%%%%%%%%%%%%%%%%%%%%%%%%%%%%%%%%%%%%%%%%%%%%%%%%%%%%%%%%%%%%%%%%%%%%%
\section{$\g{O}(D,D)$ spinors and the $\g{Pin}(D,D)$ and $\g{Spin}(D,D)$ groups}
\label{S:OddSpinors}
%%%%%%%%%%%%%%%%%%%%%%%%%%%%%%%%%%%%%%%%%%%%%%%%%%%%%%%%%%%%%%%%%%%%%%%%%%%%%%%%%
We begin by reviewing the structure of spinor representations of the split signature orthogonal group $\g{O}(D,D)$. The material in here largely follows \cite{Hohm:2011zr,Hohm:2011dv} and adopts almost all of their notation. We will try to avoid repeating too much of that material and focus instead on adding a few additional details, which will be crucial for the structure of the spinorial vielbein and the connection to bispinor representations.

\subsection{Clifford algebra of $\g{O}(D,D)$ $\Gamma$-matrices}
Understanding the structure of $\g{O}(D,D)$ spinors (as with all spinors) begins with a discussion of gamma matrices $\Gamma^M$. These satisfy the basic Clifford algebra
$\{\Gamma^M, \Gamma^N\} = 2 \,\eta^{MN}$
where $\eta^{MN}$ is the invariant metric of $\g{O}(D,D)$. The Clifford algebra $\Clif(D,D)$ consists of all products of the $\Gamma$-matrices, combined with the unit element $\mathbf 1$. A complete basis is
\begin{align}
\Clif(D,D) = \textrm{span}\, (\{\mathbf 1~, \, \Gamma^M~, \, \Gamma^{MN}~, \, \cdots~, \, \Gamma^{M_1 \cdots M_{2D}} \})~.
\end{align}
For convenience, denote $\Clif_r(D,D)$ as the real vector space spanned by $\Gamma^{M_1 \cdots M_r}$. Several of these are special:
\begin{itemize}
\item $\Clif_0(D,D)$ is the only space that commutes with $\Gamma^M$: it is isomorphic to $\mathbb R$.

\item $\Clif_1(D,D)$ is isomorphic to the space of vectors in $\mathbb R^{2D}$, and alternatively provides the defining relation for the Clifford algebra
by requiring for all $V = V^M \Gamma_M \in \Clif_1(D,D)$,
$V \cdot V = \eta(V,V) \,\mathbf 1$, where $\eta(V,V) = V^M V_M$.

\item $\Clif_2(D,D)$ is isomorphic to $\mathfrak{spin}(D,D)$.

\item $\Clif_{2D}(D,D)$ consists of a single element $\Gamma^{M_1 \cdots M_{2D}}$ that \emph{anti-commutes} with $\Gamma^M$ and squares to $\mathbf 1$. This is just the chirality matrix $\Gamma_*$, which must have eigenvalues $\pm1$. With it, we define the $\g{SO}(D,D)$ invariant antisymmetric tensor as
\begin{align}\label{E:defEpsilonSODD}
\Gamma^{M_1 \cdots M_{2D}} = \veps^{M_1 \cdots M_{2D}} \, \Gamma_*~.
\end{align}
One can show that $\veps^{M_1 \cdots M_{2D}}$ obeys
\begin{align}
\veps^{M_1 \cdots M_{2D}} \veps_{N_1 \cdots N_{2D}} = (-1)^D (2D)!\, \delta_{N_1 \cdots N_{2D}}{}^{M_1 \cdots M_{2D}}~.
\end{align}
\end{itemize}

The $\Gamma^M$ matrices have a convenient realization in the toroidal basis of $\g{O}(D,D)$, where
\begin{align}\label{E:etaCurved}
\eta^{MN} =
\begin{pmatrix}
0 & \delta^m{}_n \\
\delta_m{}^n & 0
\end{pmatrix}
\end{align}
Here they can be identified with fermionic raising and lowering operators $\psi^m$ and $\psi_m$,
\begin{align}
\Gamma^M &= (\Gamma^m, \Gamma_m) = \frac{1}{\sqrt 2} (\psi^m, \psi_m) \quad \implies \quad \eol
\{\psi^m, \psi^n\} &= \{\psi_m, \psi_n\} = 0~, \qquad
\{\psi^m, \psi_n\} = \delta_n{}^m~.
\end{align}
Interpreting this as an oscillator algebra, we may define a Clifford vacuum $\ket{0}$ as a state annihilated by all the $\psi_m$. It will be convenient to take the Clifford vacuum to have positive chirality, i.e. $\Gamma_* \ket{0} = \ket{0}$. This resolves a sign ambiguity in \eqref{E:defEpsilonSODD}, fixing
\begin{align}\label{E:defGamma*}
\Gamma_* &= (\psi^0 + \psi_0) \cdots (\psi^{D-1} + \psi_{D-1}) 
    \times (\psi^{D-1} - \psi_{D-1}) \cdots (\psi^0 - \psi_0) \eol
        &= (\psi_0 \psi^0 - \psi^0 \psi_0) \cdots (\psi_{D-1} \psi^{D-1} - \psi^{D-1} \psi_{D-1})
\end{align}
with $\veps^{M_1 \cdots M_{2D}}$ given by $\veps_0{}^0{}_1{}^1\cdots{}_{D-1}{}^{D-1} = +1$.
The element $\Gamma_*$ is $(-1)^{N_F}$ of \cite{Hohm:2011zr,Hohm:2011dv}.

\subsection{Additional bases of $\Gamma$ matrices}
It will be useful to give two other bases of $\Gamma$ matrices. The first is
\begin{align}
\Gamma_\pm^m = \psi^m \pm \psi_m~. 
\end{align}
These can be thought of as the $\Gamma$ matrices in the basis where 
$\eta^{MN} = \mathrm{diag}(\delta^{mn}, -\delta^{mn})$, as they obey
\begin{align}\label{E:Gamma_pm}
\{\Gamma_\pm^m, \Gamma_\pm^n\} = \pm 2 \delta^{mn}~, \quad
\{\Gamma_+^m, \Gamma_-^n\} = 0~.
\end{align}
These are useful when discussing the maximal \emph{compact} subalgebra, $\g{SO}(D) \times \g{SO}(D)$. 

Another useful basis is
\begin{align}\label{E:GammaLRm}
\Gamma_L^m = \psi^m + \eta^{mn} \psi_n~, \qquad
\Gamma_R^m = \psi^m - \eta^{mn} \psi_n~,
\end{align}
where $\eta^{mn} = \textrm{diag}(-1,+1,\cdots,+1)$. These are useful for the double Lorentz subalgebra $\g{SO}(D-1,1) \times \g{SO}(1,D-1)$ and are related to the previous set by
\begin{align}
\Gamma_L^0 &= \Gamma_-^0~, \qquad
\Gamma_R^0 = \Gamma_+^0~, 
\eol
\Gamma_L^k &= \Gamma_+^k~,
\qquad
\Gamma_R^k = \Gamma_-^k~, \qquad k=1,2,\cdots,D-1~.
\end{align}
We will usually write the latter with an $a$ index, i.e. $\Gamma_L^a$ and $\Gamma_R^a$. Be aware that this does \emph{not} mean that they are dressed with any vielbein; rather, all $\Gamma$ matrices will be constants.

\subsection{Involutions of the Clifford algebra}
We will need several different involutions of the Clifford algebra. First, following \cite{Hohm:2011zr,Hohm:2011dv}, we introduce $\tau$ and $\star$ as two different \emph{anti-involutions}, which flip the ordering of elements. On any two elements $S$ and $T$ of the Clifford algebra, they act as
\begin{alignat}{2}
\tau(S T) &= \tau(T) \tau(S)~, &\qquad \tau(\Gamma^M) &= \Gamma^M~, \\
(S T)^\star &= T^\star S^\star~, &\qquad (\Gamma^M)^\star &= - \Gamma^M = \Gamma_* \Gamma^M \Gamma_*~.
\end{alignat}
It will be useful to give a special name to the composition of $\star$ and $\tau$. Denoting it by $\widetilde{\phantom a}$, it is an involution and acts as
\begin{align}
\widetilde{(S T)} &= \widetilde{S} \,\widetilde{T}~, \qquad \widetilde{\Gamma^M} = -\Gamma^M~.
\end{align}
An element $S$ is called \emph{even} if $\widetilde S = S$ and \emph{odd} if $\widetilde S = - S$. Not all of these operations are independent. For example,
\begin{align}
S^\star = \Gamma_* \tau(S) \Gamma_*~, \qquad
\widetilde S = \Gamma_* S \Gamma_*~.
\end{align}

Because the $\Gamma^M$ satisfy the Clifford algebra, so do their complex conjugates, transposes, and Hermitian conjugates, and these are related by similarity transformations. Typically one introduces $A$, $B$, and $C$ matrices obeying\footnote{In principle, there
can also be signs in the below formulae, but since the number of dimensions $2D$ is even, we can choose positive signs by including factors of $\Gamma_*$ in $A$, $B$, and $C$ as needed.}
\begin{align}
(\Gamma^M)^\dag = A \Gamma^M A^{-1}~, \qquad
(\Gamma^M)^* = B \Gamma^M B^{-1}~, \qquad
(\Gamma^M)^T = C \Gamma^M C^{-1}~.
\end{align}
Because Hermitian conjugation follows from composing transposition with complex conjugation, the $A$ matrix can be related to the $B$ and $C$ matrices. Moreover, if the $\Gamma$ matrices are unitary, $(\Gamma^M)^\dag = \Gamma_M$, then one can show that $A$, $B$, and $C$ must be unitary also.

In the oscillator representation, $\psi^m$ and $\psi_m = (\psi^m)^\dag$ are real operators. This implies that $\Gamma^M$ obeys
\begin{align}
(\Gamma^M)^\dag = (\Gamma^M)^T = \Gamma_M~, \qquad (\Gamma^M)^* = \Gamma^M~,
\end{align}
which is a unitary Majorana representation: the $B$ matrix is the identity, and the unitary $A$ and $C$ matrices coincide. Following the conventions of \cite{Hohm:2011zr, Hohm:2011dv}, we choose them to be\footnote{We will follow the convention of Hohm, Kwak, and Zwiebach \cite{Hohm:2011zr, Hohm:2011dv} of referring to this element as the charge conjugation matrix $C$; but it is often more naturally thought of as the $A$ matrix. For example, its definition for even $D$ is as the product of all timelike $\Gamma^M$ matrices -- exactly as the $A$ matrix is usually defined -- and for odd $D$, it is the same multiplied by $\Gamma_*$ (equivalently, the product of all spacelike $\Gamma^M$).}
\begin{align}\label{E:CMatrix}
A = C :=
\begin{cases}
(\psi^0 - \psi_0) \cdots (\psi^{D-1} - \psi_{D-1}) & \text{$D$ even} \\
(\psi^0 + \psi_0) \cdots (\psi^{D-1} + \psi_{D-1}) & \text{$D$ odd} 
\end{cases}~.
\end{align}
From these, one can show that indeed
\begin{align}
C \psi_m C^{-1} = \psi^m~, \qquad
C \psi^m C^{-1} = \psi_m \quad \implies \quad
C \Gamma^M C^{-1} = (\Gamma^M)^\dag = \Gamma_M~.
\end{align}
Hermitian conjugation is yet another anti-involution, obeying $(S T)^\dag = T^\dag S^\dag$. It is related to $\tau$ via
\begin{align}
S^\dag = C \tau(S) C^{-1}~.
\end{align}
There are several other useful relations involving $C$:
\begin{align}
C^{-1} &= C^\dag = (-1)^{D (D-1)/2} C~, \qquad
C^\star = (-1)^D C^{-1}~, \qquad
C \Gamma_* C^{-1} = (-1)^D \Gamma_*~.
\end{align}
Additional useful relations involving $\Gamma_*$ are
\begin{align}
\tau(\Gamma_*) = (\Gamma_*)^\star = (-1)^D \Gamma_*~, \qquad \widetilde\Gamma_* = \Gamma_*~, \qquad
(\Gamma_*)^\dag = \Gamma_*~.
\end{align}

\subsection{The $\g{Pin}(D,D)$ and $\g{Spin}(D,D)$ groups}
In order to define spinors, we first introduce the groups $\g{Pin}(D,D)$ and $\g{Spin}(D,D)$.
The group $\g{Pin}(D,D)$ is defined, as in \cite{Hohm:2011zr,Hohm:2011dv}, by
\begin{align}\label{E:DefPin}
\g{Pin}(D,D) :=
\Big\{
\Lambda \in \Clif(D,D) \;\; &| \;\;
\Lambda \cdot V \cdot \Lambda^{-1} \in \Clif_1(D,D) \;\; \forall V \in \Clif_1(D,D)~,
\eol & \qquad 
\Lambda \cdot \Lambda^\star = \pm \mathbf 1
\Big\}~.
\end{align}
Let's briefly explain this definition.
The first condition in \eqref{E:DefPin}, that $\Lambda$ takes vectors into vectors, is equivalent to saying $\Lambda$ acts on the $\Gamma$-matrices as\footnote{This 
convention for $\rho(\Lambda)_N{}^M$ is the only
major deviation from \cite{Hohm:2011zr,Hohm:2011dv}.}
\begin{align}\label{E:LambdaGammaLambda}
\Lambda \Gamma^M \Lambda^{-1} = \Gamma^N \rho(\Lambda)_N{}^M 
\end{align}
for some $\rho(\Lambda)_N{}^M$.
The function $\rho(\Lambda)$ is clearly a homomorphism,
$\rho(\Lambda_1 \Lambda_2) = \rho(\Lambda_1) \rho(\Lambda_2)$, and obeys
\begin{align}
\rho(\tilde \Lambda) = \rho(\Lambda)~, \qquad
\rho(x \Lambda) = \rho(\Lambda) ~, \,\,\forall x\in \mathbb R \setminus \{0\}~.
\end{align}
Moreover, if $\rho(\Lambda) = 1$, then $\Lambda$ must commute with $\Gamma^M$,
implying that $\Lambda \in \Clif_0(D,D) = \mathbb R$.
The element $\rho(\Lambda)_N{}^M$ must also be an element in $\g{O}(D,D)$.
To see this, observe for any $V^M$ that we may define
\begin{align}
V' := \Gamma^M \rho(\Lambda)_M{}^N V_N = \Lambda V \Lambda^{-1}~.
\end{align}
It follows that
\begin{align}
\eta(V', V') \,\mathbf 1 
    = \eta(\Lambda V \Lambda^{-1}, \Lambda V \Lambda^{-1}) \,\mathbf 1
    = \Lambda V V \Lambda^{-1} = \eta(V,V) \mathbf 1~.
\end{align}
The second condition in \eqref{E:DefPin} is just a normalization. A straightforward
exercise shows that $\rho(\Lambda) = \rho((\Lambda^\star)^{-1}) = \rho(\Lambda^\star)^{-1}$ and so
$\rho(\Lambda \cdot \Lambda^\star) = 1$. Thus $\Lambda \cdot \Lambda^\star$ must be an element
of $\Clif_0(D,D)$, and so we can normalize it to $\pm \mathbf 1$.

This establishes that to every element in $\g{Pin}(D,D)$, one can associate an element of
$\g{O}(D,D)$. This mapping is $2:1$ since one can show that of the normalized elements,
precisely $\Lambda$ and $-\Lambda$ map to the same $\rho(\Lambda)_N{}^M$.
One can further show that $\rho$ is a surjective map by providing a constructive
definition for every element $\rho(\Lambda)_M{}^N\in \g{O}(D,D)$ the explicit elements
$\pm\Lambda$ that map to it. This construction can be found in \cite{Hohm:2011zr,Hohm:2011dv}, and we do not repeat it here. This establishes that $\g{Pin}(D,D)$ is a double cover of $\g{O}(D,D)$.

There are four disconnected components of $\g{O}(D,D)$, which we denote by
$\g{O}^{(\alpha,\beta)}(D,D)$ in terms of two signs $\alpha$ and $\beta$. These correspond 
to possible orientation reversals in the \emph{compact} $\g{O}(D) \times \g{O}(D)$ 
subgroups,\footnote{Note the distinction with the alternative non-compact subgroup
$\g{O}(D-1,1)_L \times \g{O}(1, D-1)_R$.} and these lead to four disconnected components of $\g{Pin}(D,D)$.
Because $\rho(\widetilde\Lambda) = \rho(\Lambda)$ implies $\widetilde\Lambda = \pm \Lambda$, $\Lambda$ must be purely even or odd.
It follows that $\tau(\Lambda) = \pm \Lambda^\star$, and so
$\Lambda \Lambda^\star = \pm 1$ is equivalent to
$\Lambda \tau(\Lambda) = \pm 1$. The various possibilities
are characterized by two signs $\alpha$ and $\beta$:\footnote{The differences between $D$ even and odd in \eqref{E:def.alphabeta} arise because
$\tau(\Gamma_*) = (\Gamma_*)^\star = (-1)^D \Gamma_*$.}
\begin{align}\label{E:def.alphabeta}
\Lambda \Lambda^\star = 
\begin{cases}
\alpha_\Lambda & \text{$D$ even} \\
\beta_\Lambda & \text{$D$ odd} 
\end{cases}~, \qquad
\Lambda \tau(\Lambda) =
\begin{cases}
\beta_\Lambda  & \text{$D$ even} \\
\alpha_\Lambda & \text{$D$ odd} 
\end{cases}~, \qquad
\tilde\Lambda = \alpha_\Lambda \beta_\Lambda \Lambda~.
\end{align}
These correspond to the four disconnected pieces $\g{Pin}^{(\alpha,\beta)}(D,D)$.
For each of these four disconnected components, we give a characteristic element and its spinor analogue:
\begin{alignat}{3}\label{E:Pin_elements}
\mathbf 1 &\in \g{Pin}^{(+,+)}(D,D) &\qquad 
\rho(\mathbf 1) &=
\begin{pmatrix}
1 & 0 \\
0 & 1
\end{pmatrix} &&\in \g{O}^{(+,+)}
~, \eol
\Gamma_+^{m} \Gamma_* &\in \g{Pin}^{(-,+)}(D,D) &\qquad 
\rho(\Gamma_+^{m} \Gamma_*)  &=
\begin{pmatrix}
k_m & 0 \\
0 & 1
\end{pmatrix} &&\in \g{O}^{(-,+)}
~, \eol
\Gamma_-^{m} \Gamma_* &\in \g{Pin}^{(+,-)}(D,D) &\qquad 
\rho(\Gamma_-^{m} \Gamma_*)  &=
\begin{pmatrix}
1 & 0 \\
0 & k_m
\end{pmatrix} &&\in \g{O}^{(+,-)}
~, \eol
\Gamma_+^{m} \Gamma_-^{m} &\in \g{Pin}^{(-,-)}(D,D) &\qquad 
\rho(\Gamma_+^m \Gamma_-^{m})  &=
\begin{pmatrix}
k_m & 0 \\
0 & k_m
\end{pmatrix} &&\in \g{O}^{(-,-)}
\end{alignat}
where $k_m = \textrm{diag}(1,\cdots,-1,\cdots 1)$ has a $-1$ in the $m$ slot.
We have written $\rho_M{}^N$ in the same diagonal basis where
$\eta^{MN} = \textrm{diag}(\delta^{mn}, - \delta^{mn})$.
In the original toroidal basis, these $\rho_M{}^N$ would act on a vector
$x^M = (x^m, \tilde x_m)$ as
\begin{alignat}{2}
\rho(\Gamma^m_+ \Gamma_*) &: \quad (x^m, \tilde x_m) \rightarrow -(\tilde x_m, x^m)~, &\qquad
&\phantom{=} \text{other $x$ unchanged}~, \eol
\rho(\Gamma^m_- \Gamma_*) &: \quad (x^m, \tilde x_m) \rightarrow +(\tilde x_m, x^m)~, &\qquad
&\phantom{=} \text{other $x$ unchanged}~, \eol
\rho(\Gamma^m_+ \Gamma^m_-) &: \quad (x^m, \tilde x_m) \rightarrow - (x^m, \tilde x_m)~, &\qquad
&\phantom{=} \text{other $x$ unchanged}~.
\end{alignat}
This means that $\Gamma^m_- \Gamma_* = (\psi^m - \psi_m) \Gamma_*$ acts as
a conventional $T$-duality in the $m$ direction, 
exchanging the coordinate $x^m$ and the dual winding coordinate $\tilde x_m$.
Note that $\Gamma_*$ itself corresponds to a complete inversion,
$\rho(\Gamma_*) = -1$; this lies in $\g{O}^{(+,+)}$ if $D$ is even and in $\g{O}^{(-,-)}$ if $D$ is odd.

Of the four disconnected components, only $\g{O}^{(+,+)}(D,D)$ and $\g{Pin}^{(+,+)}(D,D)$
constitute groups. This is because any element of $\g{Pin}(D,D)$ squares to an element
of $\g{Pin}^{(+,+)}(D,D)$. This in turn implies that any element of the disconnected
components $\g{Pin}^{(\alpha,\beta)}(D,D)$ can be written as a characteristic element
times an element of $\g{Pin}^{(+,+)}(D,D)$.

We define $\g{Spin}(D,D)$ so that $\Lambda$ must be even, i.e. $\alpha_\Lambda=\beta_\Lambda$,
\begin{align}\label{E:DefSpin}
\g{Spin}(D,D) :=
\{ \Lambda \in \g{Pin}(D,D) \;\; | \;\; \Lambda = \tilde \Lambda \}~.
\end{align}
It is clear that this group is itself disconnected,
\begin{align}
\g{Spin}(D,D) = \g{Pin}^{(+,+)}(D,D) + \g{Pin}^{(-,-)}(D,D)~.
\end{align}
Its two connected parts can be denoted $\g{Spin}^\pm(D,D) := \g{Pin}^{(\pm,\pm)}(D,D)$.

From the above considerations, it is obvious that elements of $\g{Spin}(D,D)$ 
should correspond to the double cover of $\g{SO}(D,D)$. 
It is useful to demonstrate this explicitly by checking whether the determinant of $\rho(\Lambda)_M{}^N$ is positive or negative for $\Lambda \in \g{Spin}(D,D)$.
Observe that
\begin{align}
\det \rho_M{}^N
    &= \frac{(-1)^D}{(2D)!}  \veps^{M_1 \cdots M_{2D}} \veps_{N_1 \cdots N_{2D}} 
        \rho_{M_1}{}^{N_1} \cdots\rho_{M_{2D}}{}^{N_{2D}}~.
\end{align}
Now using the definition of $\Gamma^*$ and using $\vert_{\Clif_0}$ to denote
projection to the coefficient of $\mathbf 1$,
\begin{align}
\det \rho_M{}^N
%         &= \frac{(-1)^D }{(2D)!} \,
%         \Gamma^{M_1 \cdots M_{2D}} \cdot \Gamma_* \cdot \Gamma_{N_1 \cdots N_{2D}} \cdot \Gamma_* \,
%         \rho_{M_1}{}^{N_1} \cdots\rho_{M_{2D}}{}^{N_{2D}} \Big\vert_{\Clif_0} \eol
    &= \frac{(-1)^D}{(2D)!}  \,
        \Gamma^{M_1 \cdots M_{2D}} \cdot \Gamma_{N_1 \cdots N_{2D}}\,
        \rho_{M_1}{}^{N_1} \cdots\rho_{M_{2D}}{}^{N_{2D}} \Big\vert_{\Clif_0}\eol
    &= \frac{(-1)^D}{(2D)!} \,
        \Gamma^{M_1 \cdots M_{2D}} 
            \cdot \Lambda^{-1} \cdot \Gamma_{M_1 \cdots M_{2D}} \cdot \Lambda \Big\vert_{\Clif_0}\eol
    &= \Gamma_* \cdot \Lambda^{-1} \cdot \Gamma_* \cdot \Lambda \Big\vert_{\Clif_0}
    = \tilde\Lambda^{-1} \cdot \Lambda \Big\vert_{\Clif_0} 
    = \alpha_\Lambda \beta_\Lambda~.
\end{align}
This is positive precisely when $\alpha_\Lambda$ and $\beta_\Lambda$ are both positive or both negative.
An equivalent form of the above identity, which we will need in what follows, is 
\begin{align}
\Lambda \cdot \Gamma_* \cdot \Lambda^{-1} = \Gamma_* \det \rho(\Lambda)
    = \alpha_\Lambda \beta_\Lambda \Gamma_*~.
\end{align}

Finally, there is one more identity we need. Because of \eqref{E:def.alphabeta}, 
\begin{align}
\Lambda^\dag = C \tau(\Lambda) C^{-1} =
\begin{cases}
\beta_\Lambda \,C \Lambda^{-1} C^{-1} & \text{$D$ even} \\
\alpha_\Lambda \,C \Lambda^{-1} C^{-1} & \text{$D$ odd}
\end{cases}~.
\end{align}
If $\Lambda \in \g{Spin}^\pm(D,D)$, then $\alpha_\Lambda = \beta_\Lambda = \pm 1$ and this collapses to
\begin{align}
\Lambda^\dag = \pm C \Lambda^{-1} C^{-1} \qquad
\Lambda \in \g{Spin}^\pm(D,D)~.
\end{align}

\subsection{$\g{O}(D,D)$ spinors}
We are now in a position to discuss $\g{O}(D,D)$ spinors. Given the Clifford vacuum $\ket{0}$ obeying
\begin{align}
\psi_m \ket{0} = 0~, \qquad \Gamma_* \ket{0} = \ket{0}~,
\end{align}
a spinor $\ket{\chi}$ is constructed on the Fock space by acting with all possible raising operators:
\begin{align}\label{E:ClifSpinor}
\ket{\chi} = \sum_p \frac{1}{p!} \chi_{m_1 \cdots m_p}\, \psi^{m_1} \cdots \psi^{m_p} \ket{0}~.
\end{align}
The coefficients $\chi_{m_1 \cdots m_p}$ are $p$-forms on $\mathbb R^D$. We presume these coefficients are real and this defines $\ket{\chi}$ to be a Majorana spinor. Its dimension is $2^D$ by virtue of the numerical identity
$2^D = [0] + [1] + \cdots + [D]$
where $[p] = \binom{p}{D}$ is the dimension of a $p$-form in $D$ dimensions.
Spinors $\ket{\chi}$ can be even or odd under $\Gamma_*$ if they involve
only even $p$-forms or odd $p$-forms, and in this case they have dimension $2^{D-1}$.

The Hermitian conjugate of $\ket{\chi}$ is denoted by the bra $\bra{\chi}$,
\begin{align}
\bra{\chi} := \ket{\chi}^\dag = \sum_p \frac{1}{p!} \bra{0} \psi_{m_p} \cdots \psi_{m_1}  \chi_{m_1 \cdots m_p}\,
\end{align}
where we have used $\bra{0} := \ket{0}^\dag$. The Dirac conjugate is
\begin{align}
\bra{\bar \chi} &:= \bra{\chi} C
    = \sum_p \frac{1}{p!} \bra{\bar 0} \psi^{m_p} \cdots \psi^{m_1} \chi_{m_1 \cdots m_p}~, \qquad
\bra{\bar 0} = \bra{0} C = \bra{0} \psi_1 \cdots \psi_D~.
\end{align}
The vacuum state is normalized so that $\braket{0}{0} = 1$.

The spinor $\ket{\chi}$ transforms under $\g{Pin}(D,D)$ as
$\ket{\chi'} = \Lambda \ket{\chi}$.
It suffices to understand the infinitesimal case, which generates $\g{Pin}^{(+,+)}(D,D)$, and the three non-trivial characteristic elements in \eqref{E:Pin_elements}. An infinitesimal transformation of $\mathfrak{spin}(D,D)$ acts as
\begin{align}
\delta_\lambda \ket{\chi} &= \frac{1}{4} \lambda_{MN} \Gamma^{MN} \ket{\chi} \eol
    &= \frac{1}{2} \lambda_{mn} \psi^m \psi^n \ket{\chi}
    + \lambda_m{}^n \psi^m \psi_n \ket{\chi}
    - \frac{1}{2} \lambda_p{}^p \ket{\chi}
    + \frac{1}{2} \lambda^{mn} \psi_m \psi_n \ket{\chi}
\end{align}
from which one can read off the transformation of the various constituents.
The first term corresponds to a shift of the polyform $\chi$ by $\lambda_2 \wedge \chi$
where $\lambda_{(2)} = \frac{1}{2} \lambda_{mn} \rd x^m \wedge \rd x^n$. The second term corresponds to a $\g{GL}(D)$ rotation with parameter $\lambda_m{}^n$, and the third
term is a $\g{GL}(D)$ weight factor.
The fourth term involves the double interior product with a bivector 
$\lambda^{(2)} = \frac{1}{2} \lambda^{mn} \pa_m \wedge \pa_n$.

For the discrete transformations, we find that for $\Gamma^m_- \Gamma_*$
with $n_i$ denoting indices not equal to $m$,
\begin{align}
\chi'_{n_1 \cdots n_p} = \chi_{n_1 \cdots n_p m}~, \qquad
\chi'_{n_1 \cdots n_p m} = \chi_{n_1 \cdots n_p}~.
\end{align}
In other words, the index $m$ is deleted or added from the right of the $p$-forms.
For $\Gamma^m_+ \Gamma_*$, one finds instead
\begin{align}
\chi'_{n_1 \cdots n_p} = -\chi_{n_1 \cdots n_p m}~, \qquad
\chi'_{n_1 \cdots n_p m} = \chi_{n_1 \cdots n_p}
\end{align}
so the deletion of the index $m$ is associated with a sign flip.
As a check, these two operations square respectively to $+1$ and $-1$.
Finally, for $\Gamma^m_+ \Gamma^m_-$, one finds the presence of an $m$ index
leads to a sign flip.

Meanwhile, the Dirac conjugate of $\ket{\chi}$ transforms as
\begin{align}
\bra{\bar{\chi}'} := \ket{\chi'}^\dag C = \bra{\bar{\chi}} C^{-1} \Lambda^\dag C
    = \bra{\bar{\chi}} \tau(\Lambda)
= \begin{cases}
\beta_\Lambda\, \bra{\bar\chi} \Lambda^{-1} & \text{$D$ even} \\
\alpha_\Lambda\, \bra{\bar\chi} \Lambda^{-1} & \text{$D$ odd}
\end{cases}
\end{align}
This means that $\braket{\bar\chi'}{\chi}$ is an $\g{SO}^+(D,D)$ invariant, but
not in general an $\g{O}(D,D)$ invariant.

\subsection{The Ramond-Ramond spinor and its field strength}

In double field theory, the Ramond-Ramond sector is encoded in an $\g{O}(D,D)$ spinor. This spinor was denoted $\ket{\chi}$ in \cite{Hohm:2011zr, Hohm:2011dv}. We will reserve that notation for a generic $\g{O}(D,D)$ spinor and use $\ket{C}$ when we are speaking specifically about the Ramond-Ramond spinor potential. The RR spinor potential $\ket{C}$ has a local gauge invariance and a field strength $\ket {F}$ given similarly as
\begin{align}
\delta_\lambda \ket{C} = \slashed{\pa} \ket{\lambda}~, \qquad
\ket{F} = \slashed{\pa} \ket{C} ~, \qquad \slashed{\pa} := \psi^M \pa_M~.
\end{align}
Due to the section condition, $\slashed{\pa}^2 = 0$ and $\ket{F}$ is gauge invariant. It also satisfies the Bianchi identity $\slashed{\pa}\ket{F} = 0$.

Under $\g{O}(D,D)$ diffeomorphisms, a weight $w$ spinor $\ket{\chi}$ and its Dirac conjugate $\bra{\bar\chi}$ transform as\footnote{The normalization of the $\pa_{[M} \xi_{N]}$ term is fixed by requiring that the infinitesimal $\mathfrak{spin}(D,D)$ transformation match the infinitesimal $\mathfrak{so}(D,D)$ transformation of vectors.}
\begin{align}
\delta_\xi \ket{\chi} &= 
    \xi^N \pa_N \ket{\chi} + \frac{1}{2} \pa_{[M} \xi_{N]} \Gamma^{MN} \ket{\chi}
    + w \,\pa_M \xi^M\, \ket{\chi} \eol
    &=
    \xi^N \pa_N \ket{\chi} + \frac{1}{2} \pa_M \xi_N \Gamma^{M} \Gamma^{N} \ket{\chi}
    + (w - \tfrac{1}{2}) \,\pa_M \xi^M\, \ket{\chi}~, \\[2ex]
\delta_\xi \bra{\bar{\chi}}
%     &= 
%     \xi^N \pa_N \bra{\bar{\chi}} 
%     + \frac{1}{2} \bra{\bar \chi} \Gamma^{N M} \pa_M \xi_N 
%     + w \,\pa_M \xi^M\, \bra{\bar \chi} \eol
    &= 
    \xi^N \pa_N \bra{\bar{\chi}} 
    + \frac{1}{2}  \bra{\bar \chi} \Gamma^{N}\Gamma^M   \pa_M \xi_N
    + (w - \tfrac{1}{2}) \,\pa_M \xi^M\, \bra{\bar \chi}~.  
\end{align}
In order for $\ket{F} = \slashed{\pa}\ket{C}$ to also be a spinor, 
$\ket{C}$ must have weight $w=1/2$. This in turn implies that $\ket{F}$ also has weight $w=1/2$.

%%%%%%%%%%%%%%%%%%%%%%%%%%%%%%%%%%%%%%%%%%%%%%%%%%%%%%%%%%%%%%%%%%%%%%%%%%%%%%%%%
\section{The DFT vielbein, its spinorial analogue, and flat RR spinors}
\label{S:FlatStuff}
%%%%%%%%%%%%%%%%%%%%%%%%%%%%%%%%%%%%%%%%%%%%%%%%%%%%%%%%%%%%%%%%%%%%%%%%%%%%%%%%%
We are now in a position to discuss the spinorial version of the double vielbein,
which can be used to convert a $\g{O}(D,D)$ spinor (or ``curved spinor'') into an $\g{O}(D-1,1) \times \g{O}(1,D-1)$ bispinor (or ``flat spinor''). In this section, we will be treating the collective $\g{O}(D-1,1) \times \g{O}(1,D-1)$ indices as a single flattened version of the $\g{O}(D,D)$ spinor index. This is for two reasons. First, it allows us to focus on the issues that arise when employing the spinorial vielbein. Second, it allows us to treat the cases where $D$ is even or odd simultaneously. The bispinor and doublet bispinor formalisms will be elaborated upon in sections \ref{S:BispinorDeven} and \ref{S:BispinorDodd}.

\subsection{The DFT vielbein and metric}
Let's begin by briefly reviewing the structure of the DFT vielbein and metric.
The double vielbein is an invertible matrix $V_M{}^A$ obeying a relation
\begin{align}\label{E:etaV}
\eta^{M N} V_M{}^A = \eta^{A B} V_B{}^N
\end{align}
in terms of the $\g{O}(D,D)$ metric $\eta$. A vielbein obeying such a relation is
called an $\g{O}(D,D)$ element, since any $\g{O}(D,D)$ element $\Lambda_M{}^N$ obeys
a similar relation $\eta^{MN} \Lambda_N{}^P = \eta^{P Q} (\Lambda^{-1})_Q{}^M$.

There is a subtlety here. The index $A$ is not really the same kind of
index as $M$, and so $\eta^{AB}$ is not necessarily the same matrix as
$\eta^{MN}$. They coincide when we take the toroidal decomposition
of tangent indices, i.e. $V^A = (V^a, V_a)$, but they differ in the chiral decomposition
$V^A = (V^\ra, V^\rba)$,
\begin{align}
\eta^{AB} =
\begin{pmatrix}
0 & \delta^a{}_b \\
\delta_a{}^b & 0
\end{pmatrix} \quad \text{or} \quad
\eta^{AB} =
\begin{pmatrix}
\eta^{\ra \rb} & 0 \\
0 & \eta^{\ol{\ra\rb}}
\end{pmatrix}~, \quad
\eta^{\ol{\ra\rb}} = - \eta^{\ra\rb}~.
\end{align}
Naturally, this is merely a matter of including a constant similarity transformation $(S_0)_M{}^A$ defining what we mean by $\eta^{A B}$,
\begin{align}
\eta^{M N} (S_0)_M{}^A (S_0)_N{}^B = \eta^{A B} \quad \implies \quad
\eta^{M N} (S_0)_M{}^A = \eta^{A B} (S_0)_B{}^N~.
\end{align}
Then in turns of $S_0$, the \emph{actual} $\g{O}(D,D)$ element is in general
\begin{align}
(S_V)_M{}^N = V_M{}^A (S_0)_A{}^N \in \g{O}(D,D)~.
\end{align}
This observation is a bit of a triviality for the double vielbein, but it will be more crucial when we discuss its spinorial cousin.

Using the DFT vielbein, one can build the DFT metric 
\begin{align}
\cH_{MN} = V_M{}^A V_N{}^B \cH_{A B} 
\end{align}
where $\cH_{AB}$ decomposes in the toroidal and chiral bases as
\begin{align}
\cH_{AB} =
\begin{pmatrix}
\eta_{ab} & 0 \\
0 & \eta^{a b}
\end{pmatrix} \quad \text{or} \quad
\cH_{AB} = 
\begin{pmatrix}
\eta_{\ra\rb} & 0 \\
0 & -\eta_{\ol{\ra\rb}}
\end{pmatrix}~.
\end{align}
This matrix is invariant only under the $\g{O}(D-1,1)_L \times \g{O}(1,D-1)_R$ subgroup of $\g{O}(D,D)$. Rewriting this in terms of $(S_V)_M{}^N$, we find
\begin{align}
\cH_{MN} = (S_V)_M{}^P (\cH_0)_{P Q} (S_V^T)^Q{}_N~, \qquad
(\cH_0)_{MN} =
\begin{pmatrix}
\eta_{mn} & 0 \\
0 & \eta^{mn}
\end{pmatrix}~.
\end{align}
Alternatively, we can write $\cH_M{}^N = V_M{}^A \cH_A{}^B V_B{}^N$,
where in the toroidal and chiral decompositions
\begin{align}
\cH_A{}^B =
\begin{pmatrix}
0 & \eta_{a b} \\
\eta^{ab} & 0
\end{pmatrix} \quad \text{or} \quad
\cH_A{}^B =
\begin{pmatrix}
\delta_\ra{}^\rb  & 0 \\
0 & - \delta_\rba{}^\rbb
\end{pmatrix}~.
\end{align}
This decomposes as
\begin{align}
\cH_M{}^N &=
(S_V)_M{}^P (\cH_0)_P{}^Q (S_V^{-1})_Q{}^N~, \qquad
(\cH_0)_M{}^N =
\begin{pmatrix}
0 & \eta_{mn} \\
\eta^{mn} & 0
\end{pmatrix}~.
\end{align}

\subsection{The generalized spinorial vielbein}
\label{S:FlatStuff.SpinorV}
Just as we have introduced a double vielbein that can flatten a vector index of $\g{O}(D,D)$, we wish to introduce a spinorial version to flatten spinor indices. Here it will be notationally helpful to exhibit explicit spinor indices $\hmu$
rather than kets, although we will continue to use both languages throughout.
For example, the bras, kets, and $\Gamma$ matrices discussed in the previous section
now correspond to\footnote{Since the $B$-matrix is the identity, the Majorana condition means that $\chi_\hmu$ is real.}
\begin{align}
\ket{\chi} = \chi_\hmu~, \qquad 
\bra{\chi} = (\chi_\hmu)^* = \chi_\hmu~, \qquad
\Gamma^M = (\Gamma^M)_\hmu{}^\hnu~, \qquad 
\bra{\bar \chi} = \chi_\hmu C^{\hmu \hnu}~.
\end{align}
The natural spinorial analogue of $V_M{}^A$ is $V_\hmu{}^\halpha$ with inverse
$V_\halpha{}^\hmu$, obeying
\begin{align}\label{E:VspinorGammaVspinor}
V_\halpha{}^\hmu \,(\Gamma^M)_\hmu{}^\hnu \,V_{\hnu}{}^\hbeta = 
    (\Gamma^A)_\halpha{}^\hbeta V_A{}^M~.
\end{align}
Here we have explicitly exhibited curved spinor indices $\hmu$
and their flat versions $\halpha$. 
The above relation can be rewritten in bra/ket language as
\begin{align}\label{E:DefSpinorV}
\bra{V_\halpha} \Gamma^M \ket{V^\hbeta} = 
    (\Gamma^A)_\halpha{}^\hbeta V_A{}^M~.
\end{align}
with $\bra{V_\halpha}$ and $\ket{V^\hbeta}$ obeying 
\begin{align}\label{E:SpinorV.Norm}
\braket{V_\halpha}{V^\hbeta} = \delta_\halpha{}^\hbeta~, \qquad
\ket{V^\halpha} \bra{V_\halpha} = \mathbf 1~.
\end{align}

Later on, $\halpha$ will be interpreted
as a bispinor index in the case of even $D$ (see section \ref{S:BispinorDeven}) 
and as a bispinor doublet index in the case of odd $D$ (see section \ref{S:BispinorDodd}).
For now, we will keep our formula abstract since we can address common aspects of both
cases simultaneously.

As we have previously stressed regarding $\eta^{AB}$, we have not given an independent definition of what we mean by $(\Gamma^A)_\halpha{}^\hbeta$; in principle it can differ from
$(\Gamma^M)_\hmu{}^\hnu$ by a similarity transformation,
\begin{align}\label{E:FlatGammaDef}
(S_0)_\halpha{}^\hmu \,(\Gamma^M)_\hmu{}^\hnu \,(S_0)_{\hnu}{}^\hbeta = 
    (\Gamma^A)_\halpha{}^\hbeta (S_0)_A{}^M
\end{align}
where all objects above are constant.
We take \eqref{E:FlatGammaDef} as the formal \emph{definition} of the flat $\Gamma^A$
matrices. \emph{We emphasize that these are constant matrices.} Later when we specialize to $D$ even or odd, we will give an explicit representation for $\Gamma^A$ in terms of lower dimensional gamma matrices.

Now we may identify $(\mathbb S_V)_\hmu{}^\hnu := V_\hmu{}^\halpha (S_0)_\halpha{}^\hnu$
as a $\g{Pin}(D,D)$ element obeying
\begin{align}
\mathbb S_V^{-1} \Gamma^M \mathbb S_V = \Gamma^N \rho(\mathbb S_V^{-1})_N{}^M~, \qquad
\rho(\mathbb S_V^{-1})_N{}^M = (S_0)_N{}^A V_A{}^M~.
\end{align}
Just as $(S_0)_M{}^A$ can be chosen to be the identity,
the same can be true for $(S_0)_\hmu{}^\halpha$;
however, this is not necessarily the most convenient choice. We have
chosen a Majorana basis for $\Gamma^M$ and arranging for
$\Gamma^A$ to also be Majorana (when ultimately it will be given 
in terms of, say, $\g{SO}(D-1,1)_L \times \g{SO}(1,D-1)_R$ $\gamma$-matrices) may be
highly dimension-dependent. In such a scenario, the spinorial $S_0$
serves as the similarity transformation to this basis.

We have written the above expressions with explicit indices to emphasize that
$V_\halpha{}^\hmu$, like $V_A{}^M$, involves both flat and curved indices. Again, $\halpha$ is formally a flat spinor index of $\g{Spin}(D,D)$, but will be decomposed in terms of
$\g{Spin}(D-1,1)_L \times \g{Spin}(1,D-1)_R$ indices, just as the flat index $A$ is
treated as an $\g{SO}(D-1,1)_L + \g{SO}(1,D-1)_R$ index. Subsequently, we will
interpret $V_\halpha{}^\hmu$ as a bispinor-valued bra $\bra{\slashed{V}}$, but for now
it is notationally simpler to work with explicit indices.

Locally, $V_\halpha{}^\hmu$ encodes the same information as $V_A{}^M$ but provides a double cover; both $\pm V_\halpha{}^\hmu$ give the same $V_A{}^M$. Effectively, that suggests to treat $V_\halpha{}^\hmu$ as the independent field variable and not $V_A{}^M$. Recall that $V_A{}^M$ transforms under generalized diffeomorphisms and double Lorentz transformations as
\begin{align}
\delta V_A{}^M = \xi^N \pa_N V_A{}^M - V_A{}^N (\pa_N \xi^M - \pa^M \xi_N) + \Lambda_A{}^B V_B{}^M~.
\end{align}
It is natural to take $V_\halpha{}^\hmu$ to transform as a conjugate spinor of weight $w=0$
with an additional flat transformation on its $\halpha$ index, i.e.
\begin{align}
\delta V_\halpha{}^\hmu &= \xi^N \pa_N V_\halpha{}^\hmu
    + \frac{1}{2} V_\halpha{}^\hnu (\Gamma^{N M})_\hnu{}^\hmu \, \pa_M \xi_N
    + \frac{1}{4} \Lambda_{A B} (\Gamma^{A B})_\halpha{}^\hbeta V_\hbeta{}^\hmu~.
\end{align}
One can check that this transformation is consistent with \eqref{E:VspinorGammaVspinor}.

\subsection{Flat spinors and spinor field strengths}
Finally, we can discuss what happens when we flatten spinors.
Given a spinor $\ket{C} = C_\hmu$ of weight $w=1/2$, we must convert it to a scalar. 
The obvious way to do this is to define
\begin{align}
\widehat C_\halpha = e^{d} \, V_\halpha{}^\hmu C_\hmu
\end{align}
where the factor involving the DFT dilaton $d$ ensures that $\widehat C_\halpha$ has vanishing weight. The field strength $\ket{F} = F_\hmu$, also of weight $w=1/2$, should be flattened similarly,
\begin{align}\label{E:def.Falpha}
\widehat F_\halpha = e^d \, V_\halpha{}^\hmu F_\hmu~.
\end{align}
What relations can we surmise between $\widehat F_\halpha$ and $\widehat C_\halpha$?

The original field strength and potential are related by
\begin{align}
F_\hmu = \frac{1}{\sqrt 2} (\Gamma^M)_\hmu{}^\hnu \pa_M C_\hmu~.
\end{align}
We may freely replace $\pa_M$ with $\cD_M$ (which carries the double Lorentz connection)
as $\ket{C}$ is a Lorentz scalar. Subsequently flattening with the vielbein
and dilaton, we get
\begin{align}
\sqrt 2\, \widehat F_\halpha := \sqrt{2} \,e^{d} V_\halpha{}^\hmu F_\hmu 
    &= (\Gamma^A)_\halpha{}^\hgamma \cD_A \widehat C_\hgamma
    + (\Gamma^A)_\halpha{}^\hbeta \widehat J_{A\hbeta}{}^\hgamma \widehat C_\hgamma
\end{align}
where
\begin{align}
\widehat J_{A \hbeta}{}^\hgamma
    = - \cD_A d\,\,\delta_\hbeta{}^\hgamma
    + \frac{1}{4} J_{A B C} (\Gamma^{B C})_\hbeta{}^\hgamma~, \qquad
J_{ABC} = -\cD_A V_B{}^M \cV_{M C}~.
\end{align}
It is crucial here that while $V_\hmu{}^\halpha$ is a double cover of $V_M{}^A$,
any small variation $\delta V_\hmu{}^\halpha = J_\hmu{}^\hnu V_\hnu{}^\halpha$ 
is valued in the Lie algebra of $\mathfrak{spin}(D,D)$, and can be directly related
to the $\mathfrak{so}(D,D)$-valued variation $J_M{}^N$ of $V_M{}^A$. It follows that
(see e.g. an analogous discussion in \cite{Coimbra:2011nw})
\begin{align}
\sqrt{2} \widehat F_\halpha
    &= (\Gamma^A)_\halpha{}^\hbeta \cD_A \widehat C_\hbeta
    + \frac{1}{12} T_{A B C} (\Gamma^{A B C})_\halpha{}^\hbeta \widehat C_\hbeta 
    + \frac{1}{2} T_A\, (\Gamma^A)_\halpha{}^\hbeta \widehat C_\hbeta 
\end{align}
in terms of the DFT torsions $T_{ABC}$ and $T_A$, which are typically constrained
to vanish as a conventional constraint on the double Lorentz connection. This recovers the covariantized expression for the curvature $\widehat F_\halpha$,
\begin{align}\label{E:Falpha.DCalpha}
\widehat F_\halpha = \frac{1}{\sqrt 2} (\Gamma^A)_\halpha{}^\hbeta \cD_A \widehat C_\hbeta~.
\end{align}
Analogous formulae follow for the gauge transformation of $\widehat C_\halpha$ and the Bianchi identity on $\widehat F_\halpha$ as they are structurally identical.

\subsection{Similarity transformations and flattened spinorial objects}
Let's elaborate a bit on the similarity transformation \eqref{E:FlatGammaDef}. We can write it as
\begin{align}
(\Gamma^A)_\halpha{}^\hbeta =
    (S_0)_\halpha{}^\hmu \,\Big(\Gamma^M (S_0)_M{}^A\Big){}_\hmu{}^\hnu \,(S_0)_{\hnu}{}^\hbeta~.
\end{align}
If we take the index $A$ to be in the chiral basis and $(S_0)_M{}^A$ to be the similarity transformation to that basis, the above expression corresponds to
\begin{align}\label{E:FlatGammaDef.v2}
(\Gamma^\ra)_\halpha{}^\hbeta = (S_0^{-1} \Gamma_L^a S_0)_\halpha{}^\hbeta~, \qquad
(\Gamma^\rba)_\halpha{}^\hbeta = (S_0^{-1} \Gamma_R^a S_0)_\halpha{}^\hbeta
\end{align}
where we use \eqref{E:GammaLRm}.
It follows that the $A$, $B$, $C$, and $\Gamma_*$ matrices are
\begin{subequations}
\begin{alignat}{2}
C^{\halpha \hbeta} &= (S_0^T)^\halpha{}_\hmu C^{\hmu \hnu} (S_0)_\hnu{}^\hbeta
    &\,
    &= \Big(S_0^T C S_0\Big){}^{\halpha \hbeta}~, \\
A^{\halpha \hbeta} &= (S_0^\dag)^\halpha{}_\hmu C^{\hmu \hnu} (S_0)_\hnu{}^\hbeta
    &\,
    &= \Big(S_0^\dag C S_0\Big){}^{\halpha \hbeta}~, \\
B_\halpha{}^\hbeta &= (S_0^*{}^{-1})_\halpha{}^\hmu \delta_\hmu{}^\hnu (S_0)_\hnu{}^\hbeta
    &\,
    &= \Big(S_0^*{}^{-1} S_0\Big){}_\halpha{}^{\hbeta}~, \\
(\Gamma_*)_\halpha{}^\hbeta &= (S_0^{-1})_\halpha{}^\hmu (\Gamma_*)_\hmu{}^\hnu 
    (S_0)_\hnu{}^\hbeta
    &\,
    &= \Big(S_0^{-1} \Gamma_* S_0\Big){}^{\halpha \hbeta}
\end{alignat}
\end{subequations}
Because the last condition is in the form of a similarity transformation, $(\Gamma_*)_{\halpha}{}^\hbeta$ will take the same form, written in flat indices, as it took with curved indices \eqref{E:defGamma*},
\begin{align}
(\Gamma_*)_\halpha{}^\hbeta 
    &= \Big(S_0^{-1}\Gamma_R^0 \Gamma_L^{1 \cdots (D-1)} \Gamma_R^{(D-1) \cdots 1} \Gamma_L^0 S_0 \Big){}_\halpha{}^\hbeta \eol
    &= \Big(\Gamma^{\ol 0} \Gamma^{1 \cdots (D-1)} \Gamma^{\ol{(D-1) \cdots 1}} \Gamma^0 \Big){}_\halpha{}^\hbeta
\end{align}
In general, this will not be the case with $A$, $B$, or $C$. An exception is when $\Gamma^A$ is unitary, $(\Gamma^A)^\dag = \Gamma_A$. Then the similarity transformation $S_0$ is also unitary, and then the $A$ matrix is structurally the same as the old $A$ matrix, i.e.
\begin{align}
A^{\halpha \hbeta} = \Big(S_0^{-1} C S_0\Big){}_\halpha{}^{\hbeta}
= \begin{cases}
(\Gamma^0 \Gamma^{\ol{1\cdots (D-1)}})_\halpha{}^\hbeta & \text{$D$ even} \\
(\Gamma^{\ol 0} \Gamma^{1 \cdots (D-1)})_\halpha{}^\hbeta & \text{$D$ odd} 
\end{cases}
\end{align}

Care must be taken with these formulae, because experience with general relativity teaches one to flatten indices with the vielbein. \emph{We do not do this with spinorial indices on the $\Gamma$ matrices and other related objects.} For example, flattening $\Gamma_*$ with the spinorial vielbein leads to a sign difference with our definition of the flat $\Gamma_*$,
\begin{align}\label{E:Gamma*Flatsign}
V_\halpha{}^\hmu (\Gamma_*)_\hmu{}^\hnu V_\hnu{}^\hbeta
    = (S_0)_\halpha{}^\hmu
        (\mathbb S_V^{-1} \Gamma_* \mathbb S_V)_\hmu{}^\hnu (S_0)_\hnu{}^\hbeta
    = \alpha_V \beta_V \times (\Gamma_*)_\halpha{}^\hbeta~.
\end{align}
The factors $\alpha_V$ and $\beta_V$ are the signs corresponding to the $\g{O}^{(\alpha,\beta)}(D,D)$ component to which the spinorial vielbein belongs. Similarly,
\begin{align}\label{E:CCurvedToFlat}
(V^T)^\halpha{}_\hmu C^{\hmu \hnu} V_\hnu{}^\hbeta
    &= (S_0^T){}^\halpha{}_\hmu \Big(\mathbb S_V^\dag C \mathbb S_V \Big)^{\hmu \hnu}
        (S_0)_\hnu{}^\hbeta
    = (S_0^T){}^\halpha{}_\hmu  \Big(C \tau(\mathbb S_V) \mathbb S_V \Big)^{\hmu \hnu}
        (S_0)_\hnu{}^\hbeta \eol
&= C^{\halpha \hbeta} \times
\begin{cases}
\beta_V & \text{$D$ even} \\
\alpha_V & \text{$D$ odd}
\end{cases}
\end{align}
and
\begin{align}
(V^\dag)^\halpha{}_\hmu A^{\hmu \hnu} V_\hnu{}^\hbeta
&= A^{\halpha \hbeta} \times
\begin{cases}
\beta_V & \text{$D$ even} \\
\alpha_V & \text{$D$ odd}
\end{cases}~.
\end{align}

The fact that the left and right-hand sides of these expressions differ only by signs may seem a minor point, but these signs have consequences. Let us highlight the situation with $\Gamma_*$. The Ramond-Ramond field strength $\ket{F}$, viewed as an $\g{O}(D,D)$ spinor, can have positive or negative chirality depending on whether it involves even or odd rank $p$-forms. We denote this as
\begin{align}
\Gamma_* \ket{F} =
\begin{cases}
- \ket{F} & \text{IIB/IIB}^* \\
+ \ket{F} & \text{IIA/IIA}^*
\end{cases}
\end{align}
However, within the bispinor formulation of DFT \cite{Jeon:2012hp}, one typically chooses a convention where the bispinor field strength has \emph{fixed chirality}, and this is \emph{regardless of whether it involves even or odd $p$-forms}. This comes about precisely because of the numerical factor in \eqref{E:Gamma*Flatsign}. Explicitly,
\begin{align}
(\Gamma_*)_\halpha{}^\hbeta \widehat F_\hbeta =
\begin{cases}
- \alpha_V \beta_V \widehat F_\halpha & \text{IIB/IIB}^* \\
+ \alpha_V \beta_V \widehat F_\halpha & \text{IIA/IIA}^*
\end{cases}
\end{align}
Following \cite{Jeon:2012hp}, we will fix the above sign to be $-1$. This means that
\begin{align}\label{E:FlatFchirality}
(\Gamma_*)_\halpha{}^\hbeta \widehat F_\hbeta = - \widehat F_\halpha 
\quad \implies \quad
\alpha_V  \beta_V =
\begin{cases}
+ 1 & \text{IIB/IIB}^* \\
- 1 & \text{IIA/IIA}^*
\end{cases}
\end{align}
If the spinorial vielbein $V_\hmu{}^\halpha$ is an element of 
$\g{Pin}^{(\pm,\pm)}(D,D)$, we are dealing with a IIB/IIB$^*$ duality frame, and if it
is an element of $\g{Pin}^{(\pm,\mp)}(D,D)$, we are dealing with IIA/IIA$^*$. Later on in section \ref{S:TypeIIsugras} we will discuss further how to distinguish between the starred and unstarred cases and relate this prescription to that in \cite{Jeon:2012hp}.

\subsection{The Ramond-Ramond action and the spinorial metric}
The Ramond-Ramond action has been constructed in both spinor \cite{Hohm:2011zr, Hohm:2011dv} and bispinor forms \cite{Jeon:2012kd, Jeon:2012hp}. Here let us explain how to reproduce these results. As we have already labored to produce scalar field strengths $\widehat F_\halpha$, we can build a Lagrangian\footnote{The factor of the dilaton is so that the result is a scalar density.}
\begin{align}
\cL = \frac{1}{4} e^{-2d} \, \widehat F_\halpha C^{\halpha \hbeta} \cK_\hbeta{}^{\hgamma} \widehat F_\hgamma
\end{align}
given some double Lorentz-covariant object $\cK_\hbeta{}^\hgamma$.
Such an object cannot involve the spinorial vielbein because any factors of the spinorial vielbein, if covariant, would inevitably involve a contraction of their curved spinor indices, leading to an identity matrix. So we are left with constant elements of the Clifford algebra, written with flat indices. Whatever we write down, it must commute at the very least with the connected part of $\g{SO}(D-1,1)_L \times \g{SO}(1,D-1)_R$, and this is a major restriction.
Aside from the identity and $\Gamma_*$, the only other possibilities are $\Gamma^{01\cdots(D-1)}$ and $\Gamma^{\ol{01\cdots(D-1)}}$, and these two are related by multiplying by $\Gamma_*$.

Let us give this element a name. We define it in the original Clifford algebra as\footnote{This object was denoted $\Psi_+$ in \cite{Geissbuhler:2013uka} for $D=10$ where it was similarly employed for writing down the action and self-duality conditions. It was also denoted $\Gamma^{(-)}$ in \cite{Coimbra:2011nw}.}
\begin{align}
\Gamma_{R*} =
\begin{cases}
\Gamma_R^{0\cdots D-1} & \text{$D$ even} \\
\Gamma_L^{0\cdots D-1} & \text{$D$ odd} 
\end{cases}
\end{align}
This is motivated by requiring that $\Gamma_{R*}$ \emph{anticommutes} with $\Gamma_R^a$
and \emph{commutes} with $\Gamma_L^a$. We could similarly define a $\Gamma_{L*}$ but that would just be proportional to $\Gamma_{R*} \Gamma_*$. Interestingly, $\Gamma_{R*}$ is not always Hermitian or idempotent:
\begin{align}
(\Gamma_{R*})^\dag &= - (-1)^{D (D-1)/2} \Gamma_{R*}~, \qquad
(\Gamma_{R*})^2 = - (-1)^{D (D-1)/2}~.
\end{align}
For completeness, we also give
\begin{align}
\tau(\Gamma_R) = (-1)^{D (D-1)/2} \Gamma_R~, \qquad
(\Gamma_R)^\star = (-1)^{D (D+1)/2} \Gamma_R~.
\end{align}
It will be convenient to write $C$ in terms of $\Gamma_{R*}$:
\begin{align}\label{E:C=G0G0GR*}
A = C =
\begin{cases}
\Gamma_L^0 \Gamma_R^1 \cdots \Gamma_R^D & \text{$D$ even} \\
\Gamma_R^0 \Gamma_L^1 \cdots \Gamma_L^D & \text{$D$ odd}
\end{cases} \quad 
    = \Gamma_L^0 \Gamma_R^0 \Gamma_{R*}
\end{align}
The above involved curved indices; when we pass to flat indices, $\Gamma_{R*}$ retains its structural form.

Now, a double Lorentz-invariant $\cK_\halpha{}^\hbeta$ could involve any
combination of the four elements $\mathbf 1$, $\Gamma_*$, $\Gamma_{R*}$,
and $\Gamma_{R*} \Gamma_*$.
The first two options are uninteresting, because they can be rewritten in 
curved indices to be proportional to
$\braket{\bar F}{F}$ and $\bra{\bar F}\Gamma_* \ket{F}$ and both of these
are total derivatives after imposing the section condition. This leaves
the two options $\Gamma_{R*}$ and $\Gamma_{R*} \Gamma_*$.
It is easy enough to check that they are both $\g{Pin}$ elements with (in
the toroidal and chiral bases)
\begin{alignat}{2}
\rho(\Gamma_{R*})_A{}^B
&=
\begin{pmatrix}
0 & \eta_{a b} \\
\eta^{ab} & 0
\end{pmatrix} 
& \quad \text{or} \quad
\rho(\Gamma_{R*})_A{}^B &=
\begin{pmatrix}
\delta_\ra{}^\rb & 0 \\
0 & -\delta_\rba{}^\rbb
\end{pmatrix}~, \\
\rho(\Gamma_{R*} \Gamma_*)_A{}^B
&=
\begin{pmatrix}
0 & -\eta_{a b} \\
-\eta^{ab} & 0
\end{pmatrix}
& \quad \text{or}\quad 
\rho(\Gamma_{R*} \Gamma_*)_A{}^B
&=
\begin{pmatrix}
-\delta_\ra{}^\rb & 0 \\
0 & \delta_\rba{}^\rbb
\end{pmatrix}~.
\end{alignat}
The first of these is the spinorial element corresponding to $\cH_A{}^B$, and the second maps to $-\cH_A{}^B$. Comparing with \cite{Hohm:2011zr, Hohm:2011dv}, we indeed see that $\Gamma_{R*}$ should match, in the flat space limit, precisely the flattened version of their $\cK$, which was chosen to reproduce $\cH_M{}^N$.

This suggests that correct choice to make contact with \cite{Hohm:2011zr, Hohm:2011dv} is $\Gamma_{R*}$. (This was also argued more explicitly in \cite{Geissbuhler:2013uka}.) We define
\begin{align}\label{E:RRactionCurved}
\cK = - (-1)^{D (D-1)/2} \Gamma_{R*} = (\Gamma_{R*})^{-1} = 
\begin{cases}
-\Gamma_{R*} & D=0,1,4,5,8,9,\cdots \\
\phantom{+}\Gamma_{R*} & D=2,3,6,7,10,11,\cdots
\end{cases}
\end{align}
The overall sign factor may seem a bit bizarre, but we will see in just a moment that this guarantees that it coincides with the action given in \cite{Hohm:2011zr, Hohm:2011dv}.
The putative Lagrangian is
\begin{align}\label{E:FlatAction}
\cL = \frac{1}{4} \, e^{-2d} \, \widehat F_\halpha (C (\Gamma_{R*})^{-1})^{\halpha \hbeta} \widehat F_\hbeta~.
\end{align}
Unflattening the spinor indices, this becomes
\begin{align}
\cL = \frac{1}{4} \,\bra{F} (\mathbb S_V^{-1}) ^\dag C (\Gamma_{R*})^{-1} \mathbb S_V^{-1} \ket{F}~.
\end{align}
Expanding out $C$ using \eqref{E:C=G0G0GR*} gives
\begin{align}
\cL = \frac{1}{4} \bra{F} (\mathbb S_V^{-1})^\dag \Gamma_L^0 \Gamma_R^0 \mathbb S_V^{-1} \ket{F}
    = \frac{1}{4} \bra{F} (\mathbb S_V^{-1})^\dag (\psi^0 \psi_0 - \psi_0 \psi^0) \mathbb S_V^{-1} \ket{F}~.
\end{align}
Comparing with (4.10) of \cite{Hohm:2011dv}, which reads in our conventions
\begin{align}\label{E:HohmAction}
\cL = \frac{1}{4} \bra{F} \mathbb S \ket{F} = \frac{1}{4} F_\hmu \mathbb S^{\hmu \hnu} F_\hnu~,
\end{align}
we can identify the spinorial metric $\mathbb S$ of \cite{Hohm:2011zr, Hohm:2011dv} as
\begin{align}\label{E:SpinorialMetric}
\mathbb S 
= (\mathbb S_V^{-1})^\dag \Gamma_L^0 \Gamma_R^0 \mathbb S_V^{-1}
= (\mathbb S_V^{-1})^\dag (\psi^0 \psi_0 - \psi_0 \psi^0) \mathbb S_V^{-1}~.
\end{align}
While the expression (3.14) in \cite{Hohm:2011dv} is given more explicitly in terms of a further decomposition of the spinorial vielbein in terms of $b$ and $e$, it is clear that these formulae match in the flat limit where $V_M{}^A = \delta_M{}^A$ and $\mathbb S_V = 1$.
One can explicitly check that indeed
\begin{align}
\mathbb S \Gamma^M \mathbb S^{-1} = \Gamma^N \rho(\mathbb S)_N{}^M \quad \implies \quad
\rho(\mathbb S)_N{}^M = \cH^{NM}~.
\end{align}
Similarly, one can define the curved spinor 
$\cK = C^{-1} \mathbb S = V_\hmu{}^\halpha \cK_\halpha{}^\hbeta V_\hbeta{}^\hnu$ and this obeys
\begin{align}
\cK \Gamma^M \cK^{-1} = \Gamma^N \rho(\cK)_N{}^M~, \qquad \rho(\cK)_N{}^M = \cH_N{}^M~.
\end{align}

There is something peculiar about the spinorial representative $\mathbb S$, which was noted already in \cite{Hohm:2011zr, Hohm:2011dv}. Under $\g{O}(D,D)$ transformations, $\mathbb S \rightarrow (\Lambda^{-1})^\dag \mathbb S \Lambda^{-1}$. For simplicity, let $\mathbb S_V = 1$ initially so that $\mathbb S$ describes Minkowski space with vanishing Kalb-Ramond field. Under the discrete transformations \eqref{E:Pin_elements}, $\mathbb S$ flips sign under timelike (but not spacelike) T-duality:
\begin{align}
\Lambda^\dag \Gamma_L^0 \Gamma_R^0 \Lambda = 
\begin{cases}
+ \Gamma_L^0 \Gamma_R^0 
    & \Lambda \cong \Gamma_L^k \Gamma_* \text{ or } \Lambda \cong \Gamma_R^k \Gamma_* \\
- \Gamma_L^0 \Gamma_R^0 
    & \Lambda \cong \Gamma_L^0 \Gamma_* \text{ or } \Lambda \cong \Gamma_R^0 \Gamma_*
\end{cases}
\end{align}
As noted in \cite{Hohm:2011zr, Hohm:2011dv}, this is exactly what is expected because the Ramond-Ramond action should flip sign under timelike $T$-duality, which exchanges 
IIA with IIB${}^*$ and IIB with IIA${}^*$. 

In this way, within the metric formulation of DFT, the chirality of $\ket{F}$ determines whether one is in IIA/IIA${}^*$ or IIB/IIB${}^*$, and the sign convention of $\mathbb S$ determines whether one is in IIA/IIB or IIA${}^*$/IIB${}^*$. In the next section, we will describe how the bispinor formulation characterizes these various cases.

%%%%%%%%%%%%%%%%%%%%%%%%%%%%%%%%%%%%%%%%%%%%%%%%%%%%%%%%%%%%%%%%%%%%%%%%%%%%%%%%%
\section{Distinguishing type II supergravities and the double Lorentz group}
\label{S:TypeIIsugras}
%%%%%%%%%%%%%%%%%%%%%%%%%%%%%%%%%%%%%%%%%%%%%%%%%%%%%%%%%%%%%%%%%%%%%%%%%%%%%%%%%
Within the bispinor formulation, the Ramond-Ramond field strength has a fixed chirality \eqref{E:FlatFchirality} and the action takes a fixed form \eqref{E:FlatAction}. This means that the precise choice of duality frame must be encoded in the spinorial vielbein. Let us now fully explain how. Much of this material follows \cite{Jeon:2012kd,Jeon:2012hp} but the covariant connection with the $\g{O}(D,D)$ spinor formulation \cite{Hohm:2011zr, Hohm:2011dv} has, to the best of our knowledge, not been explained before.

\subsection{Parametrizing the double vielbein}
The double vielbein can be parametrized in the chiral tangent frame basis as 
(see e.g. \cite{Jeon:2012kd,Jeon:2012hp})
\begin{align}
V_M{}^A &=
\frac{1}{\sqrt 2}
\begin{pmatrix}
\delta_m{}^n & b_{mn} \\
0 & \delta^m{}_n
\end{pmatrix}
\times
\begin{pmatrix}
e_n{}^\ra & \bar e_n{}^{\rba} \\
\eta^{\ra \rb} e_\rb{}^n & \,\eta^{\ol{\ra\rb}} \bar e_{\rbb}{}^n
\end{pmatrix}~, \\
V_A{}^M &=
\frac{1}{\sqrt 2}
\begin{pmatrix}
e_\ra{}^n & \,\eta_{\ra \rb} e_{n}{}^\rb \\
\bar e_\rba{}^n & \,\eta_{\ol{\ra\rb}} \bar e_{n}{}^{\rbb}
\end{pmatrix} \times
\begin{pmatrix}
\delta_m{}^n & -b_{mn} \\
0 & \delta^m{}_n
\end{pmatrix}~,
\end{align}
where $e_m{}^\ra$ and $\bar e_m{}^\rba$ are invertible elements with inverses
$e_\ra{}^m$ and $\bar e_\rba{}^m$.\footnote{This parametrization follows simply from assuming $e_\ra{}^m := \sqrt 2 \, V_\ra{}^m $ and $e_\rba{}^m := \sqrt{2} \,V_\rba{}^m$ to both be invertible $D \times D$ matrices. This is generically true outside of a measure zero set.}
The local double Lorentz symmetries act separately on $e_m{}^\ra$ and $\bar e_m{}^\rba$ in the obvious manner.
The requirement that $V_M{}^A$ be an $\g{O}(D,D)$ element means that both vielbeins
yield the same metric:
\begin{align}
e_m{}^\ra e_n{}^\rb \eta_{\ra \rb} = -e_m{}^{\rba} e_n{}^{\rbb} \eta_{\ol{\ra \rb}} =: g_{mn}~.
\end{align}
This ensures that the double vielbein is encoding only the metric and the $b$-field. The signatures of $\eta_{\ra \rb}$ and $\eta_{\ol{\ra\rb}}$ fix the signature of the metric $g_{mn}$ as $(-+\cdots+)$.

The fact that $e_m{}^\ra$ and $\bar e_m{}^\rba$ give the same metric means that
$(e^{-1} \bar e)_\ra{}^{\rbb}$ is just a Lorentz transformation. Keeping in mind this point, it is useful to isolate the difference between $e$ and $\bar e$ into an $\g{O}(1,D-1)_R$ rotation. That is, we decompose
\begin{align}\label{E:ParkVielbein}
V_M{}^A &=
\frac{1}{\sqrt 2}
\begin{pmatrix}
\delta_m{}^n & b_{mn} \\
0 & \delta^m{}_n
\end{pmatrix}
\times
\begin{pmatrix}
e_n{}^\rb & e_n{}^{\rb} \\
\eta^{\rb \rc} e_\rc{}^n & - \eta^{\rb\rc} e_{\rc}{}^n
\end{pmatrix}
\times
\begin{pmatrix}
\delta_\rb{}^\ra & 0 \\
0 & (e^{-1} \bar e)_\rb{}^{\bar \ra}
\end{pmatrix}~.
\end{align}
We are planning to identify $e_m{}^\ra$ with the vielbein $e_m{}^a$ of supergravity,
and to identify the left-handed Lorentz group as the supergravity Lorentz group.
That leaves the right Lorentz transformation $(e^{-1} \bar e)_\ra{}^{\rbb}$ as an
explicit factor.

It was argued in \cite{Jeon:2012hp} that the determinant of $(e^{-1} \bar e)_\ra{}^{\rbb}$
determines whether one is describing IIA or IIB supergravity.
This is a statement about whether $e^{-1} \bar e$ lies in 
$\g{O}^{(\pm,\pm)}(1, D-1)$ or $\g{O}^{(\pm,\mp)}(1,D-1)$;
that is, does it involve discrete time and space reversals or not?
We will refine this point and argue that the four possibilities for $\alpha_\Lambda$
and $\beta_\Lambda$ of $\g{O}^{(\alpha_\Lambda, \beta_\Lambda)}(1,D-1)$ further differentiate between
IIA, IIA${}^*$, IIB, and IIB${}^*$. We summarize this proposal in
Table \ref{T:TypeIISugras}.

\begin{table}[t]
\begin{center}\begin{tabular}{cccc}
$S_\Lambda$ & $(\alpha, \beta)$ & Type \\ \hline
$1$ & $(+,+)$ & IIB \\
$\Gamma_R^0 \Gamma_*$ & $(-,+)$ & IIA$^*$\\
$\Gamma_R^k \Gamma_*$ & $(+,-)$ & IIA\\
$\Gamma_R^0 \Gamma_R^k$ & $(-,-)$ & IIB$^*$ 
\end{tabular}
\end{center}
\caption{Characteristic classes $\g{O}^{(\alpha,\beta)}(1,D-1)$
of $e^{-1} \bar e$. $S_\Lambda$ denotes a corresponding
discrete spinorial representative in $\g{O}^{(\alpha,\beta)}(D,D)$.
The assignment for IIB is conventional; the others follow from
T-duality, with $k$ denoting any spacelike direction.}
\label{T:TypeIISugras}
\end{table}

\subsection{A brief digression on classifying $V_A{}^M$}
Let us make a brief aside at this point about how to classify the DFT vielbein.
We have previously mentioned that any $\g{O}(D,D)$ element (such as $V_M{}^A$)
is characterized by two signs $\alpha$ and $\beta$,
corresponding to whether the element involves an orientation reversal in either
of the compact subgroups $\g{O}(D) \times \g{O}(D)$. These two signs divide the connected
components of $\g{O}(D,D)$. However, once we choose a duality frame and identify the
physical coordinates $x^m$, the set of connected pieces of $V_A{}^M$ bifurcates further.
This is because we now require that 
$V_\ra{}^m = \frac{1}{\sqrt 2} e_\ra{}^m$ and 
$V_{\rba}{}^m = \frac{1}{\sqrt 2} e_\rba{}^m$ are invertible. 

To elucidate this matter, we follow a similar argument as in 
\cite{Hohm:2011dv} and exhibit a one-parameter family of DFT vielbeins that
connects two physically disconnected cases. Let us consider a continuous $\g{O}(D,D)$
transformation that rotates a spacelike direction (say the (1) direction)
in $\g{O}(D-1,1)$ into the temporal direction of
$\g{O}(1,D-1)$. Both of these directions are in the same compact $\g{O}(D)$ subgroup,
and so they exhibit as a simple planar rotation between the 1 and $\bar 0$ indices.
Explicitly, we have
\begin{alignat}{2}
V_0{}^m (\theta) &= V_0{}^m~,
&\quad
V_m{}^0(\theta) &= V_m{}^0~, \eol
V_1{}^m(\theta) &= \cos \theta \,V_1{}^m + \sin \theta\, V_{\bar 0}{}^m~,
&\qquad
V_m{}^1(\theta) &= V_m{}^1\, \cos\theta + V_{m}{}^{\bar 0} \,\sin\theta~,
\eol
V_{\bar 0}{}^m(\theta) &= \cos \theta \,V_{\bar 0}{}^m - \sin \theta\, V_{1}{}^m ~, 
&\qquad
V_m{}^{\bar 0}(\theta) &= V_m{}^{\bar 0}\, \cos\theta - V_{m}{}^{1} \,\sin\theta~,
\eol
V_{\bar 1}{}^m(\theta) &= V_{\bar 1}{}^m
&\qquad
V_m{}^{\bar 1}(\theta) &= V_m{}^{\bar 1}~,
\end{alignat}
with the other components of $V_M{}^A$ unchanged. 
This rotation does not affect which continuous part of $\g{O}(D,D)$ the double vielbein is a member of. For definiteness, let's take the initial value of $V_A{}^M$ to be the identity
and specialize to the $D=2$ case for simplicity. Then one can show that the left and right vielbeins are
\begin{align}
e_\ra{}^m(\theta) &=
\begin{pmatrix}
1 & 0 \\
\sin\theta \,& \cos \theta
\end{pmatrix}~, \qquad
e_\rba{}^m(\theta) =
\begin{pmatrix}
\cos\theta \,& -\sin\theta \\
0 & 1
\end{pmatrix}~.
\end{align}
At $\theta=0$ both vielbeins are the identity.
At $\theta=\pi$ we find $e_\ra{}^m(\pi) = \textrm{diag}(1,-1)$ and
$e_\rba{}^m(\pi) = \textrm{diag}(-1,1)$. This means that the determinant of both
vielbeins has flipped sign. In order for this to occur, we must pass through
an unphysical (non-invertible) pair of vielbeins, in this case at $\theta=\pi/2$. 

Note that the metric and Kalb-Ramond field corresponding to this case are
\begin{align}
g^{mn}(\theta) &=
\begin{pmatrix}
-\cos^2\theta \,& \tfrac{1}{2} \sin 2\theta \\
\tfrac{1}{2} \sin 2\theta & \cos^2 \theta
\end{pmatrix}~, \qquad
b_{mn}(\theta) =
\begin{pmatrix}
0 & \tan \theta \\
-\tan\theta & 0
\end{pmatrix}~.
\end{align}
This is the same one-parameter family considered in the appendix of \cite{Hohm:2011dv}, 
where the generalized metric version of this discussion was given. There the physicality
requirement was that $\cH^{mn} = g^{mn}$ remain invertible (the same as here),
and the same one-parameter trajectory was exhibited.

The above discussion suggests that once the physical coordinates are chosen,
we may use $e = \det e_m{}^\ra$ and $\bar e = \det \bar e_m{}^\rba$ to
further characterize the double vielbein; that is, we can identify whether
$e_m{}^\ra$ and $\bar e_m{}^\rba$ lie in $\g{GL}^+(D)$ or $\g{GL}^-(D)$.
One further piece of data is available: the Lorentz transformation
$(e^{-1} \bar e)_\ra{}^\rbb$. For the case discussed above, it is
\begin{align}
(e^{-1} \bar e)_\ra{}^\rbb(\theta) =
\begin{pmatrix}
\sec\theta & \tan \theta \\
\tan\theta & \sec\theta
\end{pmatrix}
\end{align}
and this is equal to $\textrm{diag}(1,1)$ at $\theta=0$, diverges at $\theta=\pi/2$, and passes to $\textrm{diag}(-1,-1)$ at $\theta=\pi$. Since this is a Lorentz transformation,
it can be classified as an element of $\g{O}^{(\alpha_\Lambda,\beta_\Lambda)}(1, D-1)$, with $\alpha_\Lambda=-1$ and/or $\beta_\Lambda=-1$ in the presence of temporal and/or spatial orientation reversals.

\begin{table}[t]
\begin{center}\begin{tabular}{ccccc}\toprule
Type & $e_m{}^\ra$ & $(e^{-1} \bar e)_\ra{}^{\rbb} $ & $(\alpha_V, \beta_V)$ \\ \midrule
IIB & $\g{GL}^+(D)$ & $\g{O}^{(+,+)}$ & $(+,+)$ \\
IIA${}^*$ & $\g{GL}^+(D)$ & $\g{O}^{(-,+)}$ & $(-,+)$ \\
IIA & $\g{GL}^+(D)$ & $\g{O}^{(+,-)}$  & $(+,-)$ \\
IIB${}^*$ & $\g{GL}^+(D)$ & $\g{O}^{(-,-)}$ & $(-,-)$  \\ \midrule
IIB$'$ & $\g{GL}^-(D)$ & $\g{O}^{(+,+)}$ & $(-,-)$ \\
IIA$'{}^*$ & $\g{GL}^-(D)$ & $\g{O}^{(-,+)}$ & $(+,-)$ \\
IIA$'$ & $\g{GL}^-(D)$ & $\g{O}^{(+,-)}$  & $(-,+)$ \\
IIB$'{}^*$ & $\g{GL}^-(D)$ & $\g{O}^{(-,-)}$ & $(+,+)$ \\ \midrule
\end{tabular}
\end{center}
\caption{Classification of $V_A{}^M$ once a coordinate frame is chosen.
The third column gives $\g{O}^{(\alpha_\Lambda, \beta_\Lambda)}(1,D-1)$
for $e^{-1} \bar e$ while the fourth column gives $\g{O}^{(\alpha_V, \beta_V)}(D,D)$
for the full DFT vielbein.}
\label{T:TypeIISugras_v2}
\end{table}

Because $\alpha_\Lambda \beta_\Lambda = \bar e / e$, it is sufficient to classify $V_A{}^M$ purely by $e_m{}^\ra \in \g{GL}^\pm(D)$ and by $e^{-1} \bar e \in \g{O}^{(\alpha_\Lambda,\beta_\Lambda)}(1,D-1)$. The latter case, as we have already argued, corresponds to the four options IIB, IIA, IIB${}^*$ or IIA${}^*$. The opposite sign of the determinant of $e_m{}^\ra$ gives an additional four options: these are the primed type II supergravities, which can be understood as the parity conjugates of the previous four (since we should perform an orientation-reversing coordinate transformation on the physical vielbein). The possibility of these cases is apparent in the chirality of the fermions, which are related to the chirality of the Ramond-Ramond bispinor. In the purely bosonic metric formulation, one can see it in the sign of the self-duality condition.

Finally we mention how $\alpha_V$ and $\beta_V$ of $V_A{}^M$ are related to $\alpha_\Lambda$ and $\beta_\Lambda$ defined above:
\begin{align}
(\alpha_V, \beta_V) =
\begin{cases}
(+\alpha_\Lambda, +\beta_\Lambda) & e_m{}^\ra \in \GL^+(D) \\ 
(-\alpha_\Lambda, -\beta_\Lambda) & e_m{}^\ra \in \GL^-(D)
\end{cases}
\end{align}
This is easiest to confirm by checking all the possible characteristic elements.
We summarize all of these results in Table \ref{T:TypeIISugras_v2}.

To simplify the remaining presentation, we will from now on presume $e_m{}^\ra$ to lie in the connected component of $\g{GL}(D)$,
\begin{empheq}[box=\fbox]{align}
e_m{}^\ra \in \g{GL}^+(D) \quad \implies \quad  \det e_m{}^\ra\geq 0 \qquad \qquad \text{(henceforth)}
\end{empheq}
Because we are going to use $e_m{}^\ra$ as the supergravity vielbein, this reflects the typical situation in supergravity.

\subsection{The Ramond-Ramond field strength as a polyform}
The double vielbein in \eqref{E:ParkVielbein} is given by the product of three factors
$V_b \times V_e \times V_\Lambda$. Its spinorial version has a similar decomposition:
\begin{align}\label{E:SV.Decomp}
\mathbb S_V = \mathbb S_b \times \mathbb S_e \times \mathbb S_\Lambda
\end{align}
The first two are given explicitly by
\begin{align}
\mathbb S_b = \exp\Big(\frac{1}{2} b_{mn} \psi^{m} \psi^{n} \Big)~, \quad
\mathbb S_e = \exp\Big(a_{m}{}^{n} \psi^{m} \psi_{n} - \tfrac{1}{2} a_{m}{}^{m} \Big)
\end{align}
where we are presuming $e_m{}^a = (\exp a)_m{}^a$ to be a connected component (i.e. lying in $\g{GL}^+(D)$). The last factor $S_\Lambda$ corresponds to the $e^{-1} \bar e$ Lorentz 
transformation.\footnote{This was denoted $S_e$ in \cite{Jeon:2012hp}; we hope this notation
change will not confuse the reader.}

The flattened Ramond-Ramond field strength written as a ket is
\begin{align}
\ket{\widehat F} = 
    e^d \,\mathbb S_\Lambda^{-1} \mathbb S_e^{-1} \mathbb S_b^{-1} \ket{F}~.
\end{align}
Thinking of $F$ has a polyform, the factor of $\mathbb S_b^{-1}$ involves multiplying by a polyform $e^{-b}$,
\begin{align}
\ket{\widehat F} &= 
    e^d \,\mathbb S_\Lambda^{-1} \mathbb S_e^{-1} \ket{e^{-b} F}~.
\end{align}
Defining $\hat F_{m_1 \cdots m_p}$ as the components of $e^{-b} F$, we find
\begin{align}
\mathbb S_e^{-1} \ket{e^{-b} F}
    &= \sum_p \frac{1}{2^{p/2} p!} \widehat F_{m_1 \cdots m_p} 
    \mathbb S_e^{-1} \Gamma^{m_1 \cdots m_p} \ket{0} \eol
    &= \sum_p \frac{1}{2^{p/2} p!} \widehat F_{m_1 \cdots m_p} 
    (e^{-1})_{n_1}{}^{m_1} \cdots (e^{-1})_{n_p}{}^{m_p}
        \Gamma^{n_1 \cdots n_p} \mathbb S_e^{-1} \ket{0}~.
\end{align}
We replace the dummy indices $n_i$ with $a_i$, since these should be thought of as Lorentz indices. Furthermore, $\mathbb S_e^{-1} \ket{0} = e^{1/2} \ket{0}$ and
$\Gamma_L^a \ket{0} = \frac{1}{\sqrt 2} \Gamma^a\ket{0}$, so we get
\begin{align}
\mathbb S_e^{-1} \ket{e^{-b} F}
    &= e^{1/2} \sum_p \frac{1}{p!} \widehat F_{\ra_1 \cdots \ra_p} 
        \Gamma_L^{\ra_1 \cdots \ra_p} \ket{0}~.
\end{align}
Multiplying by $\mathbb S_\Lambda^{-1}$ and $e^d = e^{-1/2} e^{\phi}$ gives
\begin{align}
\ket{\widehat F}
    &= e^{\phi} \sum_p \frac{1}{p!} \widehat F_{a_1 \cdots a_p} 
        \Gamma_L^{a_1 \cdots a_p} \mathbb S_\Lambda^{-1} \ket{0}~.
\end{align}
In this last step we have exploited that $\mathbb S_\Lambda$ is purely a right-handed
Lorentz transformation.

Now we want to identify a new vacuum state as
\begin{align}\label{E:DefZvac}
\ket{Z} := \mathbb S_\Lambda^{-1} \ket{0}~.
\end{align}
Because $\ket{0}$ obeys $\Gamma_L^a \ket{0} = \Gamma_R^a \ket{0}$,
we find the modified vacuum obeys
$\Gamma_L^\ra \ket{Z} = \Gamma_R^\rbb \ket{Z} \times (\bar e^{-1} e)_{\rbb}{}^\ra$.
This is perhaps more transparently written as the condition
\begin{align}\label{E:GammaFlatZ}
\Gamma_L^\ra \ket{Z} e_\ra{}^m = \Gamma_R^\rba \ket{Z} e_\rba{}^m~.
\end{align}
This new vacuum state obeys
\begin{align}\label{E:flatVacChirality}
\Gamma_* \ket{Z} =
\begin{cases}
+ \ket{Z} & \text{IIB / IIB}^* \\
- \ket{Z} & \text{IIA / IIA}^*
\end{cases}
\end{align}
Pushing the similarity transformation $(S_0)_\halpha{}^\hmu$ through leads to the bispinor expression
\begin{align}
\widehat F_\halpha = e^{\phi} \sum_p \frac{1}{p!} \widehat F_{\ra_1 \cdots \ra_p}
        (\Gamma_L^{\ra_1 \cdots \ra_p})_\halpha{}^\hbeta Z_\hbeta
\end{align}
This provides the dictionary between curved and flattened versions of the Ramond-Ramond
spinor. As promised, $\widehat F_\halpha$ is indeed negative chirality for both IIB and IIA due to the variable vacuum chirality \eqref{E:flatVacChirality}.
We should emphasize that $\widehat F_{\ra_1 \cdots \ra_p}$ transforms only under
the left Lorentz group, while $\widehat F_\halpha$ and $Z_\halpha$ transform under both
left and right Lorentz groups.

\subsection{The Ramond-Ramond action and spinorial metric revisited}
As a final exercise, let us compute the action in flat spinor indices. We have given it as
\begin{align}
\cL = \frac{1}{4} e^{-2d} \widehat F_\halpha (C \cK)^{\halpha \hbeta} \widehat F_\hbeta~.
\end{align}
Rewriting this as a ket gives
\begin{align}
\cL = \frac{1}{4} e^{-2d} \times \bra{\widehat F} C (\Gamma_{R*})^{-1} \ket{\widehat F}
\end{align}
with
\begin{align}
\ket{\widehat F} &= e^{\phi} \sum_p \frac{1}{p!} \widehat F_{\ra_1 \cdots \ra_p}
        \Gamma_L^{\ra_1 \cdots \ra_p} \ket{Z}~, \qquad
\bra{\widehat F} = e^{\phi} \sum_p \frac{1}{p!} \widehat F_{\ra_1 \cdots \ra_p}
        \bra{Z} (\Gamma_{L}^{\ra_p})^\dag \cdots (\Gamma_L^{\ra_1})^\dag~.
\end{align}
Combining terms, we get
\begin{align}
\cL 
&= \frac{1}{4} e^{-2d} \,e^{2\phi} \sum_{p,q}
    \frac{1}{p! q! } \widehat F_{\ra_1 \cdots \ra_p} \widehat F_{\rb_1 \cdots \rb_q}
        \bra{Z} C (\Gamma_{R*})^{-1} \Gamma_{L}^{\ra_p \cdots \ra_1} \Gamma_L^{\rb_1 \cdots \rb_q} 
            \ket{Z} \eol
    &= \frac{1}{4} e \sum_{p,q}
    \frac{1}{p! q! } \widehat F_{\ra_1 \cdots \ra_p} \widehat F_{\rb_1 \cdots \rb_q}
        \bra{0} (
            \mathbb S_\Lambda^{-1})^\dag C (\Gamma_{R*})^{-1}
            \mathbb S_\Lambda^{-1} 
            \Gamma_{L}^{\ra_p \cdots \ra_1}  \Gamma_L^{\rb_1 \cdots \rb_q} 
            \ket{0}
\end{align}
Observe that
\begin{align}
(\mathbb S_\Lambda^{-1})^\dag C \Gamma_{R*} \mathbb S_\Lambda^{-1}
    &= C \mathbb S_\Lambda \Gamma_{R*} \mathbb S_\Lambda^{-1} \times
\begin{cases}
\beta_\Lambda & \text{$D$ even} \\
\alpha_\Lambda & \text{$D$ odd}
\end{cases}
\end{align}
Taking the explicit form of $\Gamma_{R*}$ gives
\begin{align}
(\mathbb S_\Lambda^{-1})^\dag C \Gamma_{R*} \mathbb S_\Lambda^{-1} &=
\begin{cases}
\beta_\Lambda C \mathbb S_\Lambda \Gamma_R^{0 \cdots (D-1)} \mathbb S_\Lambda^{-1} & \text{$D$ even} \\
\alpha_\Lambda C \mathbb S_\Lambda \Gamma_L^{0 \cdots (D-1)} \mathbb S_\Lambda^{-1} & \text{$D$ odd} 
\end{cases} \eol
&= \alpha_\Lambda C \Gamma_{R*}
\end{align}
The computation is clear for odd $D$ because $\mathbb S_\Lambda$ is purely right-handed. For even $D$, the conjugation of $\Gamma_R^{0 \cdots (D-1)}$ by $\mathbb S_\Lambda$ gives $-1$ for \emph{any} orientation-reversing transformation, which is equivalent to $\alpha_\Lambda \beta_\Lambda$; then when multiplied by $\beta_\Lambda$, this gives just $\alpha_\Lambda$ again.  Continuing, we find
\begin{align}
\cL 
    &= \frac{1}{4} \alpha_\Lambda e \sum_{p,q} 
    \frac{1}{p! q! } \widehat F_{\ra_1 \cdots \ra_p} \widehat F_{\rb_1 \cdots \rb_q}
        \bra{0} C (\Gamma_{R*})^{-1} \Gamma_{L}^{\ra_p \cdots \ra_1}  \Gamma_L^{\rb_1 \cdots \rb_q} \ket{0} \eol
    &= \frac{1}{4} \alpha_\Lambda e \sum_{p,q} 
    \frac{1}{p! q! } \widehat F_{\ra_1 \cdots \ra_p} \widehat F_{\rb_1 \cdots \rb_q}
        \bra{0} \Gamma_L^0 \Gamma_R^0 \Gamma_{L}^{\ra_p \cdots \ra_1}  \Gamma_L^{\rb_1 \cdots \rb_q} \ket{0} 
\end{align}
and we use $\bra{0} \Gamma_L^0 \Gamma_R^0 = -\bra{0}$ to get
\begin{align}
\cL 
    &= -\frac{1}{4} \alpha_\Lambda e \sum_{p} 
    \frac{1}{p!} \widehat F_{\ra_1 \cdots \ra_p} \widehat F^{\ra_1 \cdots \ra_p} \eol
    &= - \frac{1}{4} e \sum_{p} 
    \frac{1}{p!} \widehat F_{\ra_1 \cdots \ra_p} \widehat F^{\ra_1 \cdots \ra_p} \times
\begin{cases}
+1 & \text{IIA / IIB} \\
-1 & \text{IIA${}^*$ / IIB${}^*$}
\end{cases}
\end{align}
This sign is exactly what we want: the kinetic term is the standard sign for IIA and IIB
but flips sign for their starred versions.

As a consistency check, one can repeat this calculation for the type II$'$ cases. The result is the same as above but with $e$ replaced with $|e|$:
\begin{align}
\cL 
    &= - \frac{1}{4} |e| \sum_{p} 
    \frac{1}{p!} \widehat F_{\ra_1 \cdots \ra_p} \widehat F^{\ra_1 \cdots \ra_p} \times
\begin{cases}
+1 & \text{IIA$'$ / IIB$'$} \\
-1 & \text{IIA$'{}^*$ / IIB$'{}^*$}
\end{cases}
\end{align}
This ensures that the identifications made in Table \ref{T:TypeIISugras_v2} are indeed correct.

As a consistency check, we should use the explicit decomposition of the spinorial vielbein \eqref{E:SV.Decomp} to compute the spinorial metric in the action \eqref{E:HohmAction}.
Using 
\begin{align}
(\mathbb S_\Lambda^{-1})^\dag \Gamma_L^0 \Gamma_R^0 \mathbb S_\Lambda^{-1} 
    = \alpha_\Lambda \Gamma_L^0 \Gamma_R^0
\end{align}
one can show that the spinorial metric becomes
\begin{align}
\mathbb S
    = \alpha_\Lambda \, (\mathbb S_b^{-1})^\dag (\mathbb S_e^{-1})^\dag (\psi^0 \psi_0 - \psi_0 \psi^0) \mathbb S_e^{-1} \mathbb S_b^{-1}~.
\end{align}
It was shown in \cite{Hohm:2011zr, Hohm:2011dv} that this expression (without the $\alpha_\Lambda$ sign factor) gives the correct Ramond-Ramond action when using \eqref{E:HohmAction}. The sign factor $\alpha_\Lambda$ just tracks whether there has been a timelike T-duality, and this results in a flip of the overall sign.

%%%%%%%%%%%%%%%%%%%%%%%%%%%%%%%%%%%%%%%%%%%%%%%%%%%%%%%%%%%%%%%%%%%%%%%%%%%%%%%%%
\section{Bispinors in even $D$}
\label{S:BispinorDeven}
%%%%%%%%%%%%%%%%%%%%%%%%%%%%%%%%%%%%%%%%%%%%%%%%%%%%%%%%%%%%%%%%%%%%%%%%%%%%%%%%%

When $D$ is even, a spinor of $\g{SO}(D,D)$ decomposes into a bispinor of 
$\g{SO}(D-1,1)_L \times \g{SO}(1,D-1)_R$. That is, an abstract flat spinor $\widehat F_\halpha$ can be understood as a bispinor
\begin{align}
\widehat F_\halpha \longrightarrow \widehat F_\alpha{}^\balpha~, \qquad 
\slashed{\widehat F} = (F_\alpha{}^\balpha)
\end{align}
where the $\alpha$ index is a Dirac spinor index of $\g{SO}(D-1,1)_L$ and $\balpha$ is a conjugate Dirac spinor index (because it is raised) of $\g{SO}(1,D-1)_R$ (because it is barred). We will use slash notation $\slashed{\widehat F}$ to denote the matrix with entries $\widehat F_\alpha{}^\balpha$. To make this more concrete requires an explanation of the $(\Gamma^A)_\halpha{}^\hbeta$ matrices in this bispinor form.

\subsection{Bispinor description of $\Gamma^A$}
The flat $\g{SO}(D,D)$ $\Gamma^A$ matrices will be built out of
$\g{SO}(D-1,1)_L$ and $\g{SO}(1,D-1)_R$ gamma matrices. The latter are denoted
$\gamma^\ra$ and $\bar\gamma^{\rba}$ and taken to obey
\begin{align}
\{ \gamma^\ra, \gamma^\rb\} = 2\,\eta^{\ra \rb}~, \qquad
\{ \bar\gamma^\rba, \bar\gamma^\rbb\} = 2\,\eta^{\ol{\ra\rb}}~.
\end{align}
The two chirality gamma matrices $\gamma_*$ and $\bar\gamma_*$ are given by
\begin{align}\label{E:gamma*defs}
\gamma^{0 \cdots (D-1)} =
\begin{cases}
\gamma_* & D=2,6,10,\cdots \\
i \gamma_* & D = 4,8,12,\cdots
\end{cases}~, \qquad
\bar\gamma^{\ol{0 \cdots (D-1)}} =
\begin{cases}
-\bar\gamma_* & D=2,6,10,\cdots \\
i \bar\gamma_* & D = 4,8,12,\cdots
\end{cases}~.
\end{align}
The presence of an $i$ for $D=4,8,12,\cdots$ follows since we want $(\gamma_*)^2=1$. The relative signs above were chosen as a matter of later convenience; we remind the reader that all conventions are equivalent via similarity transformations. We also will need charge conjugation matrices $C_+$ and $\bar C_+$ that obey
\begin{align}
C_+ \gamma^\ra C_+^{-1} = (\gamma^\ra)^T~, \qquad
\bar C_+ \bar\gamma^\rba \bar C_+^{-1} = (\bar\gamma^\rba)^T~.
\end{align}
Since $D$ is even, one can also introduce $C_-$ and $\bar C_-$ that differ from $C_+$
and $\bar C_+$ by factors of $\gamma_*$ and $\bar \gamma_*$, and these involve a
minus sign in the above relations.

With these ingredients we can now build the flat $\Gamma^A$ matrices:
\begin{align}\label{E:FlatGammasEvenD}
\Gamma^\ra = \gamma^\ra \otimes \mathbf 1 ~, \qquad
\Gamma^\rba = \gamma_* \otimes (\bar\gamma^\rba)^T ~.
\end{align}
In more explicit notation,
\begin{subequations}
\begin{align}
(\Gamma^\ra)_\halpha{}^\hbeta &\rightarrow
    (\Gamma^\ra)_\alpha{}^\balpha{\,}^\beta{}_\bbeta = (\gamma^\ra)_\alpha{}^\beta \delta^\balpha{}_\bbeta~, \\
(\Gamma^\rba)_\halpha{}^\hbeta &\rightarrow
    (\Gamma^\rba)_\alpha{}^\balpha{\,}^\beta{}_\bbeta = 
    (\gamma_*)_\alpha{}^\beta (\bar\gamma^\rba)_\bbeta{}^\balpha~.
\end{align}
\end{subequations}
The purpose of the transposition in $\Gamma^\rba$ is so that we can flip the barred spinor indices around when acting with $\Gamma^\rba$, i.e.
\begin{align}
(\Gamma^{\rba} \cdot \widehat F)_\alpha{}^{\balpha}
    = (\gamma_*)_\alpha{}^\beta \Big((\bar\gamma^\rba){}^T\Big){}^\balpha{}_\bbeta \,\widehat F_\beta{}^\bbeta
    = (\gamma_*)_\alpha{}^\beta \widehat F_\beta{}^\bbeta (\bar\gamma^\rba)_\bbeta{}^\balpha~.
\end{align}
Then we can suppress spinor indices and simply write
\begin{align}
\Gamma^{\rba} \cdot \slashed{\widehat F} = \gamma_* \slashed{\widehat F} \bar\gamma^\rba~.
\end{align}
It is important to observe that this implies a reversal of ordering with $\Gamma^\rba$, so that e.g.
$\Gamma^\rba \Gamma^\rbb \cdot \slashed{\widehat F} = \slashed{\widehat F} \bar\gamma^\rbb \bar\gamma^\rba$.
To avoid possible ambiguity, we include a dot symbol $\cdot$ to distinguish between
$\Gamma^\rba$ viewed as acting on the collective pair of indices ${}_\alpha{}^\balpha$
and $\bar\gamma^\rba$ acting just on the $\g{SO}(1,D-1)_R$ spinor index ${}^\balpha$ from the right.

We will give the explicit form of the similarity transformation \eqref{E:FlatGammaDef} that converts the Fock space $\Gamma^M$ matrices to the flat $\Gamma^A$ matrices in
section \ref{S:BispinorDeven.ExplicitS0} below. For now, let us trust it exists and analyze the other flattened spinorial invariants of $\g{O}(D,D)$. 
The expression \eqref{E:defGamma*} for $\Gamma_*$ in terms of $\Gamma_L^a$ and $\Gamma_R^a$ holds under the similarity transformation taking us to the basis \eqref{E:FlatGammasEvenD}. This leads to the rather simple expression
\begin{align}
\Gamma_* \cdot \slashed{\widehat F}
    &= \gamma_* \slashed{\widehat F} \bar\gamma_* \qquad\implies \quad
\Gamma_* = \gamma_* \otimes (\bar\gamma_*)^T~.
\end{align}
This elegant expression for $\Gamma_*$ was one motivation for making the particular sign
choices in \eqref{E:gamma*defs}. We can also easily compute $\Gamma_{R*} = \Gamma_R^{0 \cdots (D-1)}$ to be
\begin{align}
\Gamma_{R*} &=  
\begin{cases}
\mathbf 1 \otimes (\bar\gamma_*)^T & D=2,6,10,\cdots \\
\mathbf 1 \otimes i (\bar\gamma_*)^T & D=4,8,12,\cdots 
\end{cases}
\end{align}
and the $A$ matrix to be
\begin{align}\label{E:A.bispinor}
A = \Gamma_L^0 \Gamma_R^0 \Gamma_{R*} =
\begin{cases}
\gamma^0 \gamma_* \otimes (\bar\gamma^0)^T (\bar\gamma_*)^T & D=2,6,10,\cdots \\
i \gamma^0 \gamma_* \otimes (\bar\gamma^0)^T (\bar\gamma_*)^T & D=4,8,12,\cdots
\end{cases}
\end{align}

The $\g{SO}(D,D)$ charge conjugation matrix $C$ is more problematic, because 
a similarity transformation on spinor indices does not preserve its explicit form
as a product of $\Gamma$ matrices. Instead, we must build a $C$ that accomplishes
\begin{align}\label{E:CgammaCtemp}
C \Gamma^A C^{-1} = (\Gamma^A)^T
\end{align}
for the explicit basis \eqref{E:FlatGammasEvenD}. Up to an overall normalization, it is given by
\begin{align}\label{E:FlatCevenD}
C = (C_+ \otimes \bar C_+^{-T}) \times 
\begin{cases}
-\Gamma_{R*} & D=2,6,10,\cdots \\
-i & D=4,8,12,\cdots
\end{cases}
\end{align}
The choice of signs, while looking awkward here, will ultimately give a simple expression for $\bar C_+$ in terms of $C_+$.

\subsection{The bispinor potential, field strength, and action}

The expression for the flattened Ramond-Ramond field strength in terms of a flattened potential \eqref{E:Falpha.DCalpha} can be rewritten in bispinor form as
\begin{align}
\widehat F_\halpha = \frac{1}{\sqrt 2} (\Gamma^A)_\halpha{}^\hbeta \cD_A \widehat C_\hbeta
\quad \implies \quad
\slashed{\widehat F} = \frac{1}{\sqrt 2} \gamma^\ra \cD_\ra \slashed{\widehat C}
    + \frac{1}{\sqrt 2} \gamma_* \cD_\rba \slashed{\widehat C} \bar\gamma^\rba 
    = \cD^0_+ \slashed{\widehat C}
\end{align}
where $\cD_+^0$ is a nilpotent operator acting on a bispinor $\slashed{\widehat C}$ as
\begin{align}
\cD_+^0 \slashed{\widehat C} 
    = \frac{1}{\sqrt 2} \Gamma^A \cdot \cD_A \slashed{\widehat C}
    = \frac{1}{\sqrt 2} \gamma^\ra \cD_\ra \slashed{\widehat C}
    + \frac{1}{\sqrt 2} \gamma_* \cD_\rba \slashed{\widehat C} \gamma^{\rba}~.
\end{align}
Another nilpotent operator is $\cD_-^0$ given by
\begin{align}
\cD_-^0 \slashed{\widehat C} 
    = \frac{1}{\sqrt 2} \Gamma_{R*} \Gamma^A (\Gamma_{R*})^{-1} \cdot \cD_A \slashed{\widehat C}
= \frac{1}{\sqrt 2} \gamma^\ra \cD_\ra \slashed{\widehat C}
- \frac{1}{\sqrt 2} \gamma_* \cD_\rba \slashed{\widehat C} \gamma^{\rba}~.
\end{align}
Up to a change in normalization, these operators coincide with those introduced in \cite{Jeon:2012kd}.

The bispinor form of the action follows from \eqref{E:FlatAction} and can be written as
\begin{align}
\cL = -\frac{1}{4} e^{-2d} \Tr \Big(
\bar{\slashed {\widehat F}} \slashed{\widehat F} 
\Big)
\end{align}
where we employ a dimension-dependent definition of the Dirac conjugate of a Majorana bispinor 
$\slashed{\widehat F}$:
\begin{align}
\bar{\slashed{\widehat F}} &:=
\begin{cases}
\phantom{\bar\gamma_*} \bar C_+^{-1} \slashed{\widehat F}^T C_+ \quad & D=2,6,10,\cdots\\
\bar\gamma_* \bar C_+^{-1} \slashed{\widehat F}^T C_+ \quad & D=4,8,12,\cdots
\end{cases}
\end{align}

\subsection{Explicit form of $S_0$ and the spinorial vielbein}
\label{S:BispinorDeven.ExplicitS0}

Having rewritten our abstract flat spinor expressions as bispinors, we should elucidate the explicit similarity transformation that accomplishes this. Abstractly $(S_0)_\halpha{}^\hmu$ and its inverse converted curved spinor indices to flat ones. Interpreting a flat spinor index as a bispinor and a curved spinor index as a bra or ket, we have
\begin{align}
(S_0)_\halpha{}^\hmu \rightarrow \bra{\slashed{S_0}}~, \qquad
(S_0)_\hmu{}^\halpha \rightarrow \ket{\slashed{S_0}}~.
\end{align}
The relation \eqref{E:FlatGammaDef.v2} can be rewritten
\begin{align}
\bra{\slashed{S_0}} \Gamma_L^a = \Gamma^\ra \cdot \bra{\slashed{S_0}} = \gamma^\ra \bra{\slashed{S_0}}~, \qquad
\bra{\slashed{S_0}} \Gamma_R^a = \Gamma^\rba \cdot \bra{\slashed{S_0}} = \gamma_* \bra{\slashed{S_0}} \bar\gamma^\rba~.
\end{align}
The solution to these conditions is
\begin{align}\label{E:BraS0Def}
\bra{\slashed{S_0}} = 
    \frac{1}{2^{D/4}} \sum_p \frac{1}{p!} \bra{0} \psi_{a_p} \cdots \psi_{a_1} 
        \times (\gamma^{a_1 \cdots a_p})_\alpha{}^\balpha
\end{align}
provided we relate the two sets of $\gamma$ matrices as
\begin{align}
\gamma^\ra = \gamma_* \bar\gamma^{\rba}~.
\end{align}
The inverse relation for $(S_0)_\hmu{}^\halpha$ is
\begin{align}
\ket{\slashed{S_0}} = 
    \frac{1}{2^{D/4}} \sum_p \frac{1}{p!} \psi^{a_1} \cdots \psi^{a_p} \ket{0}
        \times (\gamma_{a_p \cdots a_1})^T~.
\end{align}
The prefactors are for convenience. As a check, we can confirm
\begin{align}
(S_0)_\halpha{}^\hmu (S_0)_\hmu{}^\hbeta  = \delta_\halpha{}^\hbeta \quad \implies \quad
\frac{1}{2^{D/2}} \sum_p \frac{1}{p!} (\gamma^{a_1 \cdots a_p})_\alpha{}^\balpha
    (\gamma_{a_p \cdots a_1})_\bbeta{}^\beta =\delta_\alpha^\beta \delta^\balpha_\bbeta
\end{align}
which is the completeness relation for $\gamma$ matrices, and
\begin{align}
(S_0)_\hmu{}^\halpha (S_0)_\halpha{}^\hnu = \delta_\hmu{}^\hnu \quad \implies \quad
    \sum_p \frac{1}{p!} \psi^{a_1} \cdots \psi^{a_p} \ket{0} \bra{0} \psi_{a_p} \cdots \psi_{a_1}
    = \mathbf 1
\end{align}
which is the completeness relation for the Fock space.

The similarity transformation allows us to explicitly define the spinorial vielbein $V_\halpha{}^\hmu$. We have described it up until now as a bra $\bra{V_\halpha}$ carrying an abstract flat spinor index. Now this flat spinor index should be interpreted as a bispinor,
\begin{align}
\bra{V_\halpha} \rightarrow \bra{V_\alpha{}^\balpha}~, \qquad 
\bra{\slashed{V}} = \Big(\bra{V_\alpha{}^\balpha}\Big)
\end{align}
where we again use the slash notation to describe a bispinor as a matrix. Here $\bra{\slashed{V}}$ should be interpreted as a matrix of bras (or a matrix-valued bra).
Its defining relation \eqref{E:DefSpinorV} can be rewritten in bispinor language as
\begin{align}
\bra{\slashed{V}} \Gamma^M = 
    \gamma^\ra \bra{\slashed{V}} V_\ra{}^M +
    \gamma_* \bra{\slashed{V}} \bar\gamma^\rba V_\rba{}^M~.
\end{align}
This state is related to the similarity transformation $\bra{\slashed{S_0}}$ via
\begin{align}
\bra{\slashed{V}} = \bra{\slashed{S_0}} \mathbb S_V^{-1}~.
\end{align}
In practice, we will not need $\mathbb S_V$ and $\bra{\slashed{S_0}}$ separately, but only the spinorial vielbein $\bra{\slashed{V}}$. It is used to define the flattened field strength as in
\eqref{E:def.Falpha}, which in bispinor language is written
\begin{align}
\slashed{\widehat F} = e^d \braket{\slashed{V}}{F}~.
\end{align}

\subsection{The vacua $\ket{0}$ and $\ket{Z}$ as bispinors}
We would like to give some convention for the vacuum state $\ket{0}$, rewritten as a bispinor.
Here we need to be very careful about notation. Formally, we have to define
$0_\halpha = (S_0)_\halpha{}^\hmu 0_\hmu$
where we use the constant similarity transformation alone to define the vacuum. That means, in particular, that \emph{this vacuum is still constant}. From \eqref{E:BraS0Def}, we find
\begin{align}
\slashed{0} = \frac{1}{2^{D/4}} \delta_\alpha{}^\balpha~.
\end{align}
The matrix structure is because $\slashed{0}$ is a constant that will connect unbarred spinors to barred spinors, and our convention \eqref{E:BraS0Def} leads this to be the identity. The normalization factor correctly leads to $\braket{0}{0} = 1$.
The vacuum is also required to obey
$\Gamma_L^a \ket{0} = \Gamma_R^a \ket{0}$; as a check, we confirm that
\begin{align}\label{E:bargammaDef}
(\gamma^a)_\alpha{}^\beta \delta_\beta{}^\balpha = 
    (\gamma_*)_\alpha{}^\beta \delta_\beta{}^\bbeta (\bar\gamma^a)_\bbeta{}^\balpha
    \quad \implies \quad
    \gamma^a = \gamma_* \bar \gamma^a~.
\end{align}
In practice, we can turn this procedure around and use the bispinor form of the vacuum to determine $S_0$ and thus the convention for $\bar\gamma^\rba$ in terms of $\gamma^\ra$. Comparing this formula with \eqref{E:gamma*defs}, we find that $\bar\gamma_* = \gamma_*$, and this is yet another reason for the sign choices we made there. Similarly, when we check $\Gamma_* \ket{0} = \ket{0}$, we find $\gamma_* \bar\gamma_* = 1$, which is consistent.

Now that the bispinor vacuum is defined, we can see how the expression \eqref{E:FlatCevenD} for $C$ determines $\bar C_+$ in terms of $C_+$.  Recall that $C$ in section \ref{S:OddSpinors} obeyed
\begin{align}
\bra{0} C \psi^{m_1} \cdots \psi^{m_D} \ket{0} = 
\veps^{m_1 \cdots m_D} (-1)^{D (D-1)/2}
\end{align}
where $\veps^{0 \cdots (D-1)} = +1$. The above can be written in flattened spinor form as
\begin{align}\label{E:CalculatingC}
0_\halpha (C \Gamma_L^{a_1 \cdots a_D})^{\halpha \hbeta} 0_\hbeta
    = \veps^{a_1 \cdots a_D} (-1)^{D (D-1)/2}~.
\end{align}
Translating this to bispinor language gives for $D=2,6,10,\cdots$
\begin{align}
- \veps^{a_1 \cdots a_D} &= - \frac{1}{2^{D/2}} \Tr\Big(
    C_+ \gamma^{a_1 \cdots a_D} \bar \gamma_* \bar C_+^{-1}
\Big) = - \frac{1}{2^{D/2}} \veps^{a_1 \cdots a_D} \Tr\Big(
    C_+ \bar C_+^{-1}
\Big)
\end{align}
from which we can determine that $\bar C_+ = C_+$.
For $D=4,8,12,\cdots$ one instead finds
\begin{align}
\veps^{a_1 \cdots a_D} = - \frac{i}{2^{D/2}} \Tr\Big(
    C_+ \gamma^{a_1 \cdots a_D} \bar C_+^{-1}
\Big) = \frac{1}{2^{D/2}} \veps^{a_1 \cdots a_D} \Tr\Big(
    C_+ \gamma_* \bar C_+^{-1}
\Big) 
\end{align}
from which we can determine that $\bar C_+ = C_+ \gamma_*$. We summarize
these conditions as
\begin{align}
\bar C_+ =
\begin{cases}
C_+ & D=2,6,10,\cdots \\
C_+ \gamma_* & D=4,8,12,\cdots
\end{cases}
\end{align}
As a check, this is consistent with $\bar C_+ \bar \gamma^a \bar C_+^{-1} = (\bar\gamma^a)^T$
upon using \eqref{E:bargammaDef}. A very similar computation shows that \eqref{E:A.bispinor} is the correct expression for $A$.

The vacuum state $\ket{0}$ is not the most natural to express as a bispinor. We would really prefer to have the vacuum $\ket{Z}$ that transforms under double Lorentz transformations. The relation \eqref{E:GammaFlatZ} for this state becomes in bispinor language
\begin{align}
\gamma^\ra \slashed{Z} e_\ra{}^m = \gamma_* \slashed{Z} \bar\gamma^{\rba} \bar e_\rba{}^m
\end{align}
Using $\gamma_* \gamma^\ra = \bar\gamma^\rba$, this can be written as
\begin{align}
\bar\gamma^\rbb (e^{-1} \bar e)_\rb{}^\rba = \slashed{Z} \bar\gamma^{\rba} \slashed{Z}^{-1}
\end{align}
This implies that $\slashed{Z}$ can be interpreted as the spinor representative
for the Lorentz transformation $(e^{-1} \bar e)_\rb{}^\rba$. This makes sense because $\slashed{Z} = Z_\alpha{}^\bbeta$ is just the action of the right Lorentz transformation $\mathbb S^{-1}_\Lambda$ on the vacuum \eqref{E:DefZvac}, so that
\begin{align}
Z_\alpha{}^\bbeta = \frac{1}{2^{D/4}} \delta_\alpha{}^\balpha \Lambda_\balpha{}^\bbeta~.
\end{align}
The previous relation we found for $\bar\gamma^\rba = \gamma_* \gamma^\ra$
corresponds to $\slashed{Z} = 1$ and $(e^{-1} e)_\rb{}^\rba = \delta_\rb{}^\rba$.
As a check, we can compute
\begin{align}
\gamma^{\ra_1 \cdots \ra_D} \slashed{Z}
    = \slashed{Z} \bar\gamma^{\rbb_D \cdots \rbb_1} 
        (\bar e^{-1} e)_{\rbb_1}{}^{\ra_1} \cdots
        (\bar e^{-1} e)_{\rbb_D}{}^{\ra_D} \quad \implies \quad
\gamma_* \slashed{Z} \bar\gamma_* = \slashed{Z} \times \det(\bar e^{-1} e)~.
\end{align}
From the properties of $\g{Pin}(1,D-1)_R$, one can further show that
\begin{align}
\Lambda^T =
\begin{cases}
\alpha_\Lambda \,\bar C_+ \Lambda^{-1} \bar C_+^{-1} & D=2,6,10,\cdots \\
\beta_\Lambda \,\bar C_+ \Lambda^{-1} \bar C_+^{-1} & D=4,8,12,\cdots
\end{cases}
\end{align}
where $\alpha_\Lambda$ and $\beta_\Lambda$ correspond to the disconnected components
$\g{O}^{(\alpha_\Lambda,\beta_\Lambda)}(1,D-1)$. This expression implies
\begin{align}
\slashed{Z} \bar{\slashed{Z}} = \frac{1}{2^{D/2}} \alpha_\Lambda \mathbf 1
= \frac{1}{2^{D/2}} \times \begin{cases}
+\mathbf 1 & \text{IIA / IIB} \\
-\mathbf 1 & \text{IIA${}^*$ / IIB${}^*$}
\end{cases}
\end{align}
This result allows us to eliminate $\slashed{Z}$ from the bispinor action and reduce
it directly to a sum of squares of field strengths. Given the field strength and its conjugate,
\begin{align}
\slashed{\widehat F} = e^{\phi} \sum_p \frac{1}{p!} \widehat F_{\ra_1 \cdots \ra_p} \gamma^{\ra_1 \cdots \ra_p} \slashed{Z}~, \qquad
\bar{\slashed{\widehat F}} = e^{\phi} \sum_p \frac{1}{p!} \widehat F_{\ra_1 \cdots \ra_p} 
\bar{\slashed{Z}} \gamma^{\ra_p \cdots \ra_1} ~,
\end{align}
it is a simple exercise using cyclicity of the trace to show that
\begin{align}
\cL = -\frac{1}{4} e^{-2d} \Tr\Big( \bar{\slashed{\widehat F}} \slashed{\widehat F} \Big) 
    = -\frac{1}{4} e\,
\sum_{p} 
    \frac{1}{p!} \widehat F_{\ra_1 \cdots \ra_p} \widehat F^{\ra_1 \cdots \ra_p} \times
\begin{cases}
+1 & \text{IIA / IIB} \\
-1 & \text{IIA${}^*$ / IIB${}^*$}
\end{cases}~.
\end{align}

%%%%%%%%%%%%%%%%%%%%%%%%%%%%%%%%%%%%%%%%%%%%%%%%%%%%%%%%%%%%%%%%%%%%%%%%%%%%%%%%%
\section{Bispinor doublets in odd $D$}
\label{S:BispinorDodd}
%%%%%%%%%%%%%%%%%%%%%%%%%%%%%%%%%%%%%%%%%%%%%%%%%%%%%%%%%%%%%%%%%%%%%%%%%%%%%%%%%

When $D$ is odd, a spinor of $\g{SO}(D,D)$ decomposes into a bispinor of 
$\g{SO}(D-1,1)_L \times \g{SO}(1,D-1)_R$ times an additional doublet index.
That is, $\widehat F_\halpha$ will be understood as a doublet bispinor
\begin{align}
\widehat F_\halpha \longrightarrow \widehat F_{i\alpha}{}^\balpha~, \qquad 
\slashed{\widehat F_i} = (F_{i\alpha}{}^\balpha)
\end{align}
where the $\alpha$ index is a Dirac spinor index of $\g{SO}(D-1,1)_L$ and $\balpha$ is a conjugate Dirac spinor index of $\g{SO}(1,D-1)_R$. The index $i$ is an additional doublet index required
in order to match the dimension of the $\g{O}(D,D)$ spinor.

\subsection{Bispinor doublet description of $\Gamma^A$}
Let us work this out by explicitly building $\Gamma^\ra$ and $\Gamma^{\rba}$
out of $\g{SO}(D-1,1)_L$ and $\g{SO}(1,D-1)_R$ gamma matrices. These we take to be
$\gamma^\ra$ and $\bar \gamma^{\rba}$ as before. Because $D$ is odd, there is
no chirality gamma matrix in these subspaces, and instead we choose
\begin{align}\label{E:gammaDoddProducts}
\gamma^{0 \cdots (D-1)} =
\begin{cases}
1 & D=3,7,11,\cdots \\
i & D = 1,5,9,\cdots
\end{cases}~, \qquad
\bar\gamma^{\ol{0 \cdots (D-1)}} =
\begin{cases}
i & D=3,7,11,\cdots \\
1 & D = 1,5,9,\cdots
\end{cases}~.
\end{align}
The factors of $i$ are required for these products to square properly.
The relative sign factors of the two cases are again chosen for convenience.
Then $\Gamma^A$ and $\Gamma_*$ are given by
\begin{align}
\Gamma^\ra = \sigma_1 \otimes \gamma^\ra \otimes \mathbf 1~, \qquad
\Gamma^\rba = \sigma_2 \otimes \mathbf 1 \otimes (\bar\gamma^\rba)^T~, \qquad
\Gamma_* = \sigma_3 \otimes \mathbf 1 \otimes \mathbf 1~,
\end{align}
where we again will transpose the barred spinor index products.
We have attached an additional two-dimensional space, on which the Pauli
matrices act, to guarantee the $\Gamma^A$ Clifford algebra is satisfied.
The expression for $\Gamma_*$ follows from taking the product of all the $\Gamma^A$.
Note that the $\Gamma_*$ chirality of an $\g{O}(D,D)$ spinor is identified with chirality on the doublet space. We will give the explicit similarity transformation connecting this to $\Gamma^M$ in the next section.

The triple tensor product is a bit inconvenient to work with. As before, we can take the two $\g{Spin}$ groups to act on two sides of a bispinor. The additional doublet index can be accounted for by extending the bispinor to a bispinor doublet. That is, a field strength
$\widehat F_{i \alpha}{}^\balpha$ becomes
$\slashed{\widehat F}_i =
\begin{pmatrix}
\slashed{\widehat F}_+ \\
\slashed{\widehat F}_-
\end{pmatrix}$.
Writing the $\Gamma$ matrices as matrices on the doublet space, we have
\begin{align}
\Gamma^\ra =
\begin{pmatrix}
0 & \gamma^\ra \otimes \mathbf 1 \\
\gamma^\ra \otimes \mathbf 1 & 0
\end{pmatrix}~, \quad
\Gamma^\rba =
\begin{pmatrix}
0 & -i \mathbf 1 \otimes (\bar\gamma^\rba)^T \\
i \mathbf 1 \otimes (\bar\gamma^\rba)^T & 0
\end{pmatrix}~, \quad
\Gamma_* =
\begin{pmatrix}
\mathbf 1 & 0\\
0 & - \mathbf 1 
\end{pmatrix}~.
\end{align}
The matrix $\Gamma_{R*}$, which anticommutes with $\Gamma^\rba$ and commutes
with $\Gamma^\ra$, was defined earlier for odd $D$ as 
$\Gamma_{R*} = \Gamma_L^{0 \cdots (D-1)}$. Here that amounts to
\begin{align}
\Gamma_{R*}  &= 
\begin{cases}
\sigma_1 \otimes \mathbf 1 \otimes \mathbf 1   & D=3,7,11,\cdots \\
i \sigma_1 \otimes \mathbf 1 \otimes \mathbf 1  & D=1,5,9,\cdots 
\end{cases}
\end{align}
or in matrix form
\begin{align}
\Gamma_{R*} &=
\begin{pmatrix}
0 & \mathbf 1 \\
\mathbf 1 & 0
\end{pmatrix} \times
\begin{cases}
1 & D=3,7,11,\cdots \\
i & D=1,5,9,\cdots 
\end{cases}
\end{align}
The $A$ matrix is
\begin{align}
A = \Gamma_L^0 \Gamma_R^0 \Gamma_{R*} =
\begin{cases}
-\sigma_2 \otimes \gamma^0 \otimes (\bar\gamma^0)^T & D=3,7,11,\cdots \\
-i \sigma_2 \otimes \gamma^0 \otimes (\bar\gamma^0)^T & D=1,5,9,\cdots
\end{cases}
\end{align}
To build the charge conjugation matrix, we observe that for $\g{SO}(D-1,1)$, the charge conjugation matrix obeys
\begin{align}
(\gamma^\ra)^T = 
\begin{cases}
- C_- \gamma^\ra C_-^{-1} & D=3,7,11,\cdots \\
\phantom{+} C_+ \gamma^\ra C_+^{-1} & D=1,5,9,\cdots
\end{cases}
\end{align}
and similarly for $\bar\gamma^\rba$. Because $D$ is odd, there is no $\gamma_*$ to convert $C_+$ into $C_-$ and only one choice is available for each $D$.
In order to satisfy $C \Gamma^A C^{-1} = (\Gamma^A)^T$, $C$ must be given (up to numerical factors) by
\begin{align}\label{E:BispinorC.Dodd}
C =
\begin{cases}
-i \sigma_2 \otimes C_- \otimes (\bar C_-^{-1})^T & D=3,7,11,\cdots \\
-i \sigma_1 \otimes C_+ \otimes (\bar C_+^{-1})^T & D=1,5,7,\cdots 
\end{cases}
\end{align}
Again the precise choice of factors is made to simplify the relation between $\bar C_\pm$ and $C_\pm$.

\subsection{The vacua $\ket{0}$ and $\ket{Z}$ as doublet bispinors}
Rather than give the similarity transformation $S_0$ we will first discuss how we want the bispinor vacuum to behave. We define it as
\begin{align}
\slashed{0}_i= \frac{1}{2^{(D-1)/4}} 
\begin{pmatrix}
\mathbf 1\\
\mathbf 0
\end{pmatrix}
\end{align}
with the normalization fixed by $\braket{0}{0} = 1$.
The conditions \eqref{E:GammaFlatZ} and \eqref{E:CalculatingC} 
determine
\begin{align}
\bar \gamma^\rba = -i \gamma^\ra~, \qquad
\bar C_\pm = C_\pm~.
\end{align}
The Lorentz covariant vacuum is denoted
\begin{align}
\slashed{Z}_i = \frac{1}{2^{(D-1)/4}}
\begin{pmatrix}
\slashed{\Lambda}_+ \\
\slashed{\Lambda}_-
\end{pmatrix}~.
\end{align}
The conditions \eqref{E:GammaFlatZ} for this state become in bispinor language
\begin{align}
\gamma^\ra \slashed{Z}_i e_\ra{}^m
    = (\sigma_3)_{ij} \slashed{Z}_j \bar\gamma^\rba \bar e_\rba{}^m
\end{align}
or equivalently 
\begin{align}\label{E.Zpm.rotation.Dodd}
\slashed{\Lambda}_+ \bar\gamma^\rba (\slashed{\Lambda}_+)^{-1} &=
    + \bar\gamma^\rbb (e^{-1} \bar e)_\rb{}^\rba ~, \eol
\slashed{\Lambda}_- \bar\gamma^\rba (\slashed{\Lambda}_-)^{-1} &=
    -\bar\gamma^\rbb  (e^{-1} \bar e)_\rb{}^\rba ~.
\end{align}
We would like to interpret these relations so that $\slashed{\Lambda}_\pm$
is an element of $\g{Pin}(1,D-1)_R$. However, here we run into an interesting puzzle regarding the $\g{Pin}$ and $\g{Spin}$ groups for odd spacetime dimensions. Spinors (representations of $\g{Spin}$ groups) are \emph{not} pinors (representations of $\g{Pin}$ groups) in odd dimensions. Rather, a pinor must have double the dimension of a spinor.\footnote{See \cite{Berg:2000ne} for a pedagogical discussion.} The reason for this is one cannot effect orientation-reversing Lorentz transformations with elements of the Clifford algebra. A single gamma matrix $\gamma^a$ reflects every direction other than $a$. In even dimensions, this can be augmented with $\gamma_*$ (which flips all directions), to produce a reflection in the $a$ direction alone. In odd dimensions, there is no such remedy. So one finds that $\gamma^0$ will flip all spatial dimensions; there are an even number of these so this lies in the connected part of the group, $\g{Pin}^{(+,+)}(1,D-1)$. An element $\gamma^k$ will flip time and an odd number of spatial directions, and so this lies in $\g{Pin}^{(-,-)}(1,D-1)$. There is no way to generate an orientation-reversing element just with these elements. The only solution is to double the dimension of the Fock space by introducing a new element to the Clifford algebra that behaves just like $\gamma_*$.  The extra doublet index for $\g{Pin}(D,D)$ for odd $D$ is doing precisely this job.

Given an $\g{O}(1,D-1)$ transformation $(e^{-1} \bar e)_\rb{}^\rba$, a putative spinorial representative $\slashed\Lambda$ can satisfy at most
\begin{align}
\slashed\Lambda \bar\gamma^{\rba} \slashed\Lambda^{-1} = \bar\gamma^{\rbb} (e^{-1} \bar e)_\rb{}^\rba \times \det (e^{-1} \bar e)
\end{align}
where the extra determinant factor on the right undoes any orientation reversal. This means that depending on the orientation-reversing behavior of $e^{-1} \bar e$, $\slashed{\Lambda}$ will satisfy this relation with either a plus or a minus sign. But comparing with \eqref{E.Zpm.rotation.Dodd}, it is evident that $\slashed{\Lambda}_\pm$ corresponds to $\slashed{\Lambda}$ with the appropriate sign.
Only one of these $\slashed{\Lambda}_\pm$ is turned on in any given duality frame. We can check that this makes sense by rederiving the action of $\Gamma_*$ on $\slashed{Z}$. Starting with
\begin{align}
\gamma^{\ra_1 \cdots \ra_D} \sigma_1 \slashed{Z}
    = \sigma_2 \slashed{Z} \gamma^{\rbb_D \cdots \rbb_1} 
        (\bar e^{-1} e)_{\rbb_1}{}^{\ra_1} \cdots
        (\bar e^{-1} e)_{\rbb_D}{}^{\ra_D} 
\end{align}
one can show that
\begin{align}
(\sigma_1)_{ij} \slashed{Z}_j = -i (\sigma_2)_{ij} \slashed{Z}_j \det (e^{-1} \bar e)
\quad \implies \quad
(\sigma_3)_{ij} \slashed{Z}_j = \slashed{Z}_i \det (e^{-1} \bar e)~.
\end{align}
Then $\slashed \Lambda_+$ is turned on when $\det(e^{-1} \bar e) = +1$ and
$\slashed \Lambda_-$ is turned on when $\det(e^{-1} \bar e) = -1$.

\subsection{The similarity transformation $S_0$ and spinorial vielbein}
\label{S:BispinorDodd.ExplicitS0}
Motivated by the discussion above, we now define
\begin{align}
(S_0)_{\halpha}{}^\hmu \rightarrow \bra{\slashed{S_0}_i}~, \qquad
(S_0)_{\hmu}{}^\halpha \rightarrow \ket{\slashed{S_0}^i}
\end{align}
The relation \eqref{E:FlatGammaDef.v2}, now given as
\begin{align}
\bra{\slashed{S_0}_i} \Gamma_L^a = (\sigma_1)_{i j} \gamma^\ra \bra{\slashed{S_0}_j} ~, \qquad
\bra{\slashed{S_0}_i} \Gamma_R^a = (\sigma_2)_{i j} \bra{\slashed{S_0}_j}  \bar\gamma^{\rba}~,
\end{align}
can be solved by
\begin{align}
\bra{(S_0)_i{}_\alpha{}^\balpha}
    &= \frac{1}{2^{(D-1)/4}} \sum_p \frac{1}{p!} \bra{0} \psi_{a_p \cdots a_1} \gamma^{a_1 \cdots a_p} \otimes
    (\sigma_1)^p
    \begin{pmatrix}
    1 \\ 
    0
    \end{pmatrix} \eol
    &= \frac{1}{2^{(D-1)/4}} 
\begin{pmatrix}
\sum_{p\,{\rm even}} \frac{1}{p!} \bra{0} \psi_{a_p \cdots a_1} \gamma^{a_1 \cdots a_p} \\
\sum_{p\,{\rm odd}} \frac{1}{p!} \bra{0} \psi_{a_p \cdots a_1} \gamma^{a_1 \cdots a_p} \\
\end{pmatrix}~.
\end{align}
Its inverse is 
\begin{align}
\ket{(S_0)_{\balpha}{}^{\alpha i}}
    = \frac{1}{2^{(D-1)/4}} \sum_p \frac{1}{p!} 
    \begin{pmatrix}
    1 & 0
    \end{pmatrix} (\sigma_1)^p \otimes
    (\gamma^{a_1 \cdots a_p})_{\balpha}{}^\alpha \psi_{a_p \cdots a_1} \ket{0}~.
\end{align}
Checking the inverse property requires the completeness relations in odd $D$ of
\begin{align}
\delta_\alpha^\beta \delta^\balpha_\bbeta &= \frac{1}{2^{(D-1)/2}} \sum_{p\, {\rm even}} \frac{1}{p!} (\gamma^{a_1 \cdots a_p})_\alpha{}^\balpha
    (\gamma_{a_p \cdots a_1})_\bbeta{}^\beta \eol
    &= \frac{1}{2^{(D-1)/2}} \sum_{p\, {\rm odd}} \frac{1}{p!} (\gamma^{a_1 \cdots a_p})_\alpha{}^\balpha
    (\gamma_{a_p \cdots a_1})_\bbeta{}^\beta ~.
\end{align}

The similarity transformation again allows us to explicitly define the spinorial vielbein $V_\halpha{}^\hmu$. It is a doublet bispinor obeying
\begin{align}
\bra{\slashed{V}_i} \Gamma^M = 
    (\sigma_1)_{ij}\, \gamma^\ra \bra{\slashed{V}_j} V_\ra{}^M +
    (\sigma_2)_{ij} \,\bra{\slashed{V}_j} \bar\gamma^\rba V_\rba{}^M
\end{align}
This state is related to the similarity transformation $\bra{\slashed{S_0}}$ via
\begin{align}
\bra{\slashed{V}_i} = \bra{\slashed{S_0}_i} \mathbb S_V^{-1}~.
\end{align}
This in turn is used to build the bispinor doublet field strength (following \eqref{E:def.Falpha}) as
\begin{align}
\slashed{\widehat F}_i = e^d \braket{\slashed{V}_i}{F}
\end{align}

\subsection{The bispinor potential, field strength, and action}
The bispinor doublet field strength $\slashed{\widehat F}_i$ is given in terms of its potential $\slashed{\widehat C}_i$ as
\begin{align}
\widehat F_\halpha = \frac{1}{\sqrt 2} (\Gamma^A)_\halpha{}^\hbeta \cD_A \widehat C_\hbeta
\quad \implies \quad
\begin{pmatrix}
\slashed{\widehat F}_+ \\
\slashed{\widehat F}_-
\end{pmatrix}
= \frac{1}{\sqrt 2} 
\begin{pmatrix}
\gamma^\ra \cD_\ra \slashed{\widehat C}_- - i \cD_\rba \slashed{\widehat C}_- \bar\gamma^{\rba}
\\
\gamma^\ra \cD_\ra \slashed{\widehat C}_+ + i \cD_\rba \slashed{\widehat C}_+ \bar\gamma^{\rba}
\end{pmatrix}~.
\end{align}
One could define operators $\cD_\pm^0$ as in the even $D$ case, but they act on bispinor doublets and are more cumbersome.
The bispinor form of the action \eqref{E:FlatAction} is
\begin{align}
\cL = -\frac{1}{4} e^{-2d}\, 
\Tr \Big(
    \bar{\slashed{\widehat F}} \slashed{\widehat F}
\Big)
\end{align}
with the bispinor Dirac conjugate defined as
\begin{align}
\bar{\slashed{\widehat F}}_i :=
\begin{cases}
(\bar C_-)^{-1} \slashed{\widehat F}{}_j^T C_- (\sigma_3)_{ji} & D=3,7,11,\cdots \\
(\bar C_+)^{-1} \slashed{\widehat F}{}_i^T C_+ & D=1,5,9,\cdots
\end{cases}
\end{align}
Writing $\slashed{\widehat F}$ in terms of the covariant vacuum gives
\begin{align}
\slashed{\widehat F} &= \sum_p \frac{1}{p!}
    \widehat F_{\ra_1 \cdots \ra_p} \gamma^{\ra_1 \cdots \ra_p} (\sigma_1^p)_{ij} \slashed{Z}_j~, \\
\bar{\slashed{\widehat F}} &= \sum_p \frac{1}{p!}
    \bar{\slashed Z}_j (\sigma_1^p)_{ji} \gamma^{\ra_p \cdots \ra_1} 
    \widehat F_{\ra_1 \cdots \ra_p} 
    ~.
\end{align}
To evaluate the action further requires an identity on $\slashed{Z}_i$:
\begin{align}
\slashed{Z}_i \bar{\slashed{Z}}_j
= \frac{1}{2^{(D-1)/2}} \delta_{ij} \alpha_\Lambda \mathbf 1
= \frac{1}{2^{(D-1)/2}}  \delta_{ij} \times \begin{cases}
+\mathbf 1 & \text{IIA / IIB} \\
-\mathbf 1 & \text{IIA${}^*$ / IIB${}^*$}
\end{cases}~.
\end{align}
This does indeed hold, although its proof is less straightforward than for even $D$.
After peeling off the connected part of $\slashed{Z}_i$, one must address each of
the four discrete cases. For $D=1,5,9,\cdots$, the discrete possibilities
of $\Gamma_R^0 \Gamma_*$, $\Gamma_R^k \Gamma_*$ and $\Gamma_R^{0k}$ contribute to
$\slashed{Z}_i$ factors of $i \bar\gamma^0$, $i \bar\gamma^k$, or $\bar\gamma^k \bar\gamma^0$,
and these obey
\begin{align}
(\bar C_+)^{-1} (i \bar\gamma^0)^T \bar C_+ &= i \bar\gamma^0 = - (i \bar\gamma^0)^{-1}~, \eol
(\bar C_+)^{-1} (i \bar\gamma^k)^T \bar C_+ &= i \bar\gamma^k = + (i \bar\gamma^k)^{-1}~, \eol
(\bar C_+)^{-1} (\bar\gamma^k \bar\gamma^0)^T \bar C_+ &= \bar\gamma^0 \bar\gamma^k = 
    - (\bar\gamma^k \bar\gamma^0)^{-1}~,
\end{align}
and so an overall sign applies when a discrete time reversal is present.
The case of $D=3,7,11,\cdots$ is similar. The result is the expected expression
\begin{align}
\cL = -\frac{1}{4} e^{-2d} \Tr\Big( \bar{\slashed{\widehat F}} \slashed{\widehat F} \Big) 
    = -\frac{1}{4} e\,
\sum_{p} 
    \frac{1}{p!} \widehat F_{\ra_1 \cdots \ra_p} \widehat F^{\ra_1 \cdots \ra_p} \times
\begin{cases}
+1 & \text{IIA / IIB} \\
-1 & \text{IIA${}^*$ / IIB${}^*$}
\end{cases}
\end{align}

%%%%%%%%%%%%%%%%%%%%%%%%%%%%%%%%%%%%%%%%%%%%%%%%%%%%%%%%%%%%%%%%%%%%%%%%%%%%%%%%%
\section{Possible self-duality conditions}
\label{S:SelfDuality}
%%%%%%%%%%%%%%%%%%%%%%%%%%%%%%%%%%%%%%%%%%%%%%%%%%%%%%%%%%%%%%%%%%%%%%%%%%%%%%%%%
Suppose that we wish to impose a self-duality equation on the Ramond-Ramond field strength. This is easiest to address with flat spinors, which are manifestly $\g{O}(D,D)$-covariant.
We decompose $\widehat F_\halpha$ into left and right-handed chiral pieces, $(\widehat F_+)_\halpha$ and $(\widehat F_-)_\halpha$. Due to the chirality decomposition, the only independent double Lorentz-invariant elements of the Clifford algebra are the identity matrix and $\Gamma_{R*}$, which has even chirality for even $D$ and odd chirality for odd $D$.

If $D$ is even, the only possible form of a self-duality equation is
\begin{align}
(\widehat F_+)_\halpha = a_1 (\Gamma_{R*})_\halpha{}^\hbeta (\widehat F_+)_\hbeta~, \qquad
(\widehat F_-)_\halpha = a_2 (\Gamma_{R*})_\halpha{}^\hbeta (\widehat F_-)_\hbeta
\end{align}
for some real coefficients $a_i$. Consistency implies that $(a_i \Gamma_{R*})^2 = 1$, and so the $a_i$ are simply two independent signs, with $D$ constrained to solve $(-1)^{D (D-1)/2} = -1$. For even $D$, this means $D= 2,6,10,\cdots$.

If $D$ is odd, then the possible self-duality conditions can be written
\begin{align}\label{E:oddDselfduality}
(\widehat F_+)_\halpha = a_2 (\Gamma_{R*})_\halpha{}^\hbeta (\widehat F_-)_\hbeta~, \qquad
(\widehat F_-)_\halpha = a_1 (\Gamma_{R*})_\halpha{}^\hbeta (\widehat F_+)_\hbeta~,
\end{align}
for $a_i$ real. Consistency of these equations requires $a_1 a_2 (\Gamma_{R*})^2 = 1$,
and this is solved by choosing $a_2 = - a_1 (-1)^{D (D-1)/2}$.
Here there is no restriction on the choice of odd $D$. For 
$D=3,7,\cdots$, one finds that $a_2 = a_1$, and for
$D=1,5,\cdots$, one finds $a_2 = -a_1$.
When $D$ is odd, one cannot impose both self-duality and chirality, since \eqref{E:oddDselfduality} is non-sensical if either $\widehat F_+$ or $\widehat F_-$ vanishes.  

The possible duality conditions can be encompassed in the same general formula
\begin{align}\label{E:FlatFselfduality}
\widehat F_\halpha = 
    \Big(\frac{1}{2} (a_1 + a_2) \Gamma_{R*} + \frac{1}{2} (a_1 - a_2) \Gamma_{R*} \Gamma_*
    \Big){}_\halpha{}^\hbeta \widehat F_\hbeta
\end{align}
subject to the following conditions:
\begin{itemize}
\item $D = 2,6,10,\cdots$ $\implies$ $a_1$ and $a_2$ are independent signs,
$a_1 = \pm1$, $a_2 = \pm1$\\
There are four options: $\widehat F_\halpha = \pm (\Gamma_{R*} \widehat F)_\halpha$
and $\widehat F_\halpha = \pm (\Gamma_{R*} \Gamma_* \widehat F)_\halpha$.
\item $D = 3,7,11,\cdots$ $\implies$ $a_2 = a_1 = \pm 1$ \\
There are two options: $\widehat F_\halpha = \pm (\Gamma_{R*} \widehat F)_\halpha$.

\item $D = 1,5,9,\cdots$ $\implies$ $a_2 = -a_1 = \pm 1$ \\
There are two options: $\widehat F_\halpha = \pm (\Gamma_{R*} \Gamma_* \widehat F)_\halpha$.
\end{itemize}

Let's analyze what these conventions mean for the covariant components
$\widehat F_{\ra_1 \cdots \ra_p}$ of the field strengths.
It will be convenient to use $(\Gamma_{R*})^{-1}$ instead of $\Gamma_{R*}$ in order to eliminate a number of awkward $D$-dependent signs. Then we can show that 
\begin{align}\label{E:SelfDuality1}
\widehat F_\halpha &= \pm (\Gamma_{R*}^{-1})_\halpha{}^\hbeta \widehat F_\hbeta  \quad \implies\quad \eol
\widehat F_{\ra_1 \cdots \ra_p} 
    &= \pm (-1)^{p(p-1)/2}
    \frac{1}{(D-p)!} \veps_{\ra_1 \cdots \ra_p \rb_1 \cdots \rb_{D-p}} 
    \widehat F^{\rb_1 \cdots \rb_{D-p}} \times
\begin{cases}
\det(e^{-1} \bar e) & \text{$D$ even} \\
1 & \text{$D$ odd}
\end{cases}
\end{align}
On the other hand,
\begin{align}\label{E:SelfDuality2}
\widehat F_\halpha 
    &= \pm(\Gamma_{R*}^{-1} \Gamma_*)_\halpha{}^\hbeta \widehat F_\hbeta  \quad \implies\quad \eol
\widehat F_{\ra_1 \cdots \ra_p} &= \pm (-1)^{p(p-1)/2}
    \frac{1}{(D-p)!} \veps_{\ra_1 \cdots \ra_p \rb_1 \cdots \rb_{D-p}} 
    \widehat F^{\rb_1 \cdots \rb_{D-p}} \times
\begin{cases}
(-1)^p  & \text{$D$ even} \\
\det(e^{-1} \bar e) (-1)^p & \text{$D$ odd}
\end{cases}
\end{align}

Although we have written the self-duality condition here in flat spinor indices, it is an interesting question to connect it to the condition given in \cite{Hohm:2011zr, Hohm:2011dv}. Introducing the spinorial vielbein, the expression \eqref{E:SelfDuality1} reads in curved spinor language
\begin{align}\label{E:SelfDuality1.curved}
\ket{F} = \pm C^{-1} \mathbb S \ket{F} \times
\begin{cases}
\beta_V & \text{$D$ even} \\
\alpha_V & \text{$D$ odd} 
\end{cases}
\end{align}
where $\mathbb S$ is the spinorial metric \eqref{E:SpinorialMetric} and the $D$-dependent sign factor comes from \eqref{E:CCurvedToFlat}. The other option \eqref{E:SelfDuality2} gives
\begin{align}\label{E:SelfDuality2.curved}
\ket{F} = \pm C^{-1} \mathbb S \Gamma_* \ket{F} \times
\begin{cases}
\alpha_V & \text{$D$ even} \\
\beta_V & \text{$D$ odd} 
\end{cases}~.
\end{align}
This is an interesting result because it implies that if the self-duality condition is fixed for the flattened spinor, it carries a sign when written in terms of the curved spinor, and this sign depends on which connected $\g{Pin}$ component the spinorial vielbein belongs to.

When we fix the chirality of $\ket{F}$ via \eqref{E:FlatFchirality}, the above two options coincide up to a relative minus sign. We choose to take a plus in \eqref{E:SelfDuality1} and \eqref{E:SelfDuality1.curved} and therefore a minus in \eqref{E:SelfDuality2} and \eqref{E:SelfDuality2.curved}. 

For even $D=2,6,10,\cdots$, the chirality and self-duality conditions read in bispinor form
\begin{alignat}{3}
(\Gamma_*)_\halpha{}^\hbeta \widehat F_\hbeta &= - \widehat F_\halpha &\quad &\implies &\quad
\gamma_* \slashed{\widehat F} \bar\gamma_* &= -\slashed{\widehat F}~, \\
(\Gamma_{R*})_\halpha{}^\hbeta \widehat F_\hbeta &= \widehat F_\halpha &\quad &\implies &\quad
\slashed{\widehat F} \bar\gamma_* &= \slashed{\widehat F}
\end{alignat}
For $D=10$ IIB/IIB${}^*$, these conditions imply (for odd $p$)
\begin{align}
\widehat F_{\ra_1 \cdots \ra_p} &= + (-1)^{p(p-1)/2}
    \frac{1}{(10-p)!} \veps_{\ra_1 \cdots \ra_p \rb_1 \cdots \rb_{10-p}} 
    \widehat F^{\rb_1 \cdots \rb_{10-p}}~, \eol
&\implies \widehat F_{\ra_1 \cdots \ra_9} = 
    +\veps_{\ra_1 \cdots \ra_9 \rb_1} 
    \widehat F^{\rb_1}~, \eol
&\implies \widehat F_{\ra_1 \cdots \ra_7} = -
    \frac{1}{3!} \veps_{\ra_1 \cdots \ra_7 \rb_1 \rb_2 \rb_{3}} 
    \widehat F^{\rb_1 \rb_2 \rb_{3}}~, \eol
&\implies \widehat F_{\ra_1 \cdots \ra_5} =
    +\frac{1}{5!} \veps_{\ra_1 \cdots \ra_5 \rb_1 \cdots \rb_{5}} 
    \widehat F^{\rb_1 \cdots \rb_{5}}~.
\end{align}
For $D=10$ IIA/IIA${}^*$, these conditions imply (for even $p$)
\begin{align}
\widehat F_{\ra_1 \cdots \ra_p} &= -(-1)^{p(p-1)/2}
    \frac{1}{(10-p)!} \veps_{\ra_1 \cdots \ra_p \rb_1 \cdots \rb_{10-p}} 
    \widehat F^{\rb_1 \cdots \rb_{10-p}}~, \eol
&\implies \widehat F_{\ra_1 \cdots \ra_{10}} = 
    +\veps_{\ra_1 \cdots \ra_{10}} 
    \widehat F~, \eol
&\implies \widehat F_{\ra_1 \cdots \ra_8} = -
    \frac{1}{2!} \veps_{\ra_1 \cdots \ra_8 \rb_1 \rb_2} 
    \widehat F^{\rb_1 \rb_2 }~, \eol
&\implies \widehat F_{\ra_1 \cdots \ra_6} =
    +\frac{1}{4!} \veps_{\ra_1 \cdots \ra_6 \rb_1 \cdots \rb_{4}} 
    \widehat F^{\rb_1 \cdots \rb_{4}}~.
\end{align}

As a final comment, we elaborate on the even dimensional cases of $D=2,6,10,\cdots$, where one can further decompose a Dirac spinor $\psi_\alpha$ into a pair of Weyl spinors $(\psi_\alpha, \psi^\alpha)$ (here we repurpose the index $\alpha$ as a Weyl spinor index). The Ramond-Ramond bispinor becomes
\begin{align}
\slashed{\widehat F} =
\begin{pmatrix}
\widehat F_{\alpha}{}^\balpha & \widehat F_{\alpha\balpha} \\
\widehat F^{\alpha \balpha} & \widehat F^\alpha{}_\balpha
\end{pmatrix}~.
\end{align}
The chirality and self-duality conditions reduce this further as
\begin{alignat}{3}
\gamma_* \slashed{\widehat F} \bar\gamma_* &= -\slashed{\widehat F}
&\quad &\implies &\quad
\slashed{\widehat F} &=
\begin{pmatrix}
0 & \widehat F_{\alpha\balpha} \\
\widehat F^{\alpha \balpha} & 0
\end{pmatrix}~, \\
\slashed{\widehat F} \bar\gamma_* &= \slashed{\widehat F}
&\quad &\implies &\quad
\slashed{\widehat F} &=
\begin{pmatrix}
\widehat F_\alpha{}^{\balpha} & 0 \\
\widehat F^{\alpha \balpha} & 0
\end{pmatrix}~.
\end{alignat}
When both conditions are imposed, only $\widehat F^{\alpha\balpha}$ is present.
This bispinor, properly super-covariantized, is the one appearing  in the type II superspace torsion tensor and is the only non-vanishing on-shell component of $\slashed{\widehat F}$.

%%%%%%%%%%%%%%%%%%%%%%%%%%%%%%%%%%%%%%%%%%%%%%%%%%%%%%%%%%%%%%%%%%%%%%%%%%%%%%%%%
\section{Discussion}
\label{S:Conclusion}
%%%%%%%%%%%%%%%%%%%%%%%%%%%%%%%%%%%%%%%%%%%%%%%%%%%%%%%%%%%%%%%%%%%%%%%%%%%%%%%%%
These notes provide a (hopefully) complete and coherent discussion of the connection between the $\g{O}(D,D)$ spinor and $\g{O}(D-1,1) \times \g{O}(1,D-1)$ bispinor formulation of the Ramond-Ramond sector of double field theory. While the two different formulations have been connected before via a double-Lorentz gauge-fixing \cite{Jeon:2012kd}, we believe this is the first complete and covariant discussion. One of the interesting features we have uncovered is the covariant flat vacuum, a bispinor $\slashed{Z}$ for $D$ even and a doublet bispinor $\slashed{Z}_i$ for $D$ odd, which transforms under both the left and right Lorentz groups, and is used to construct the Ramond-Ramond polyform expansion.

There are several avenues that we have not explored. We have not considered the case of non-Riemannian backgrounds \cite{Morand:2017fnv} (see also additional works \cite{Cho:2019ofr,Park:2020ixf} on DFT and \cite{Berman:2019izh} for an extension to exceptional field theory). These can naturally be encoded in metric DFT by choosing different parametrizations of $\cH^{MN}$ which are not simply described by a pseudo-Riemannian metric and Kalb-Ramond two-form. Such backgrounds are naturally described in DFT in either the metric or vielbein formulations, and it would be interesting to discuss the connection between their Ramond-Ramond sectors.

Another application would be the construction of the Ramond-Ramond sector in type II superspace, generalizing the type I construction \cite{Butter:2021dtu}. This would be quite interesting since existing efforts in this area have focused solely on the supervielbein where the bispinor version of the Ramond-Ramond field strength is encoded \cite{Hatsuda:2014qqa}, whereas it is clear due to the work of Cederwall \cite{Cederwall:2016ukd} that the Ramond-Ramond sector should be encoded in superspinors of $\g{OSp}(D,D|2s)$. This is a work in progress, which we hope to report on in the near future.

\acknowledgments

It is a pleasure to thank Olaf Hohm, Jeong-Hyuck Park, and Ergin Sezgin for discussions, and the JHEP referee for suggesting several clarifications. This work is partially supported by the NSF under grant NSF-2112859 and the Mitchell Institute for Fundamental Physics and Astronomy at Texas A\&M University.

\bibliography{library.bib}
\bibliographystyle{utphys_mod_v2}

\end{document}